Mechanisms for anesthesia, unawareness, respiratory depression, memory replay and sleep: MHb → IPN → PAG + DRN + MRN → claustrum → cortical slow-waves.

Karin Vadovičová
Bratislava, Slovakia
*Corresponding contact: vadovick@tcd.ie
Acknowledgments: No funding, no conflict of Interests.

Abstract
My findings show what causes loss of awareness, anesthesia, memory replay, opioid induced respiratory depression (OIRD), and slow wave sleep. Opiates are fast pain relievers and anesthetics that can cause respiratory arrest. I found how mu-opioids and anesthetics by activating medial habenula (MHb) and/or interpeduncular nucleus (IPN) induce unawareness and slowdown respiration. Using DTI method I observed that human hippocampus is connected to MHb via posterior septum, while amygdala via anteromedial BNST. MHb projected to pineal gland and contralateral MHb (Vadovičová, 2014). MHb has dense mu-opioid receptors (Gardon and Faget, 2014) and strong projections to IPN. Herkenham (1981) found that MHb and IPN increase their glucose intake during anesthesia. The question is: What is the MHb-IPN circuit doing?
I found that it causes slow-wave sleep (SWS), memory replay, sharp-wave ripples, spindles, hippocampo-cortical replay of temporally, spatially and relationally bound information, serotonin-BDNF-linked growth, synaptogenesis, rest and recovery, by activating median raphe (MRN) serotonin, and by inhibiting the theta state circuit, new memories encoding, awareness, arousal, alert wakefulness, and REM sleep (Vadovičová, 2015). SWS circuit causes also natural slowdown of respiration and heart rate, while it inhibits locomotion and arousal.
This extended circuit model added role of the dentate gyrus → posterior septum → MHb → IPN → MRN → hippocampus + basal forebrain + claustrum → cortical slow-wave activity (SWA) in memory replay, ripples, loss of awareness, anesthesia, and SWS. It proposes new neural mechanism for anesthetic ketamine, nitrous oxide, and phencyclidine effects: activation of the IPN → MRN → claustrum → cortical SWA circuit by the 5-HT2a receptors in the IPN and claustrum. My brain/circuit model shows why are ketamine and psychedelics anxiolytic and antidepressant. How they by activating the 5-HT2a receptors in vACC/infralimbic cortex increase safety, well-being signal, socializing, and cognitive flexibility, and attenuate fear, worries, anger, impulsivity, self-defence, and wanting. This model claims that mu-opioids, acetylcholine, nicotine, endocannabinoids, adenosine, GLP-1RA, and substance P activate the MHb-IPN-MRN circuit which promotes rest, recovery, repair, serotonin → BDNF → proteins production, spines/synapses growth, and anti-inflammatory state.

Abbreviations:

| | |
|---|---|
| ACC | anterior cingulate cortex |
| CeA | central nucleus of amygdala |
| BAC | bed nucleus of anterior commissure |
| BNST | bed nucleus of stria terminalis |
| DG | dentate gyrus |
| DRN | dorsal raphe nucleus |
| hDBB | horizontal limb of Diagonal Band of Broca |
| IPN | interpeduncular nucleus |
| LC | locus coeruleus |
| LDT | laterodorsal tegmental nucleus |
| LH | lateral hypothalamus |
| LHb | lateral habenula |
| LPO | lateral preoptic nucleus |
| MHb | medial habenula |
| MRN | median raphe nucleus |
| NBM | nucleus basalis of Meynert |
| NI | nucleus incertus |
| NREM | non-Rapid Eye Movement (sleep) |
| NTS | nucleus tractus solitarius |
| OIRD | opiate induced respiratory depression |
| PAG | periacqueductal grey |
| PBN | parabrachial nucleus |
| PPT | pedunculopontine tegmental nucleus |
| PH | posterior hypothalamus |
| preBötC | preBötzinger complex |
| RMTg | rostromedial tegmental nucleus |
| SI | substantia innominata |
| SUM | supramamillary nucleus |
| SWA | slow-wave activity |
| SWS | slow-wave sleep |
| vACC | ventral ACC, Brodmann area 25, a25 |
| VTA | ventral tegmental area |
| ZI | zona incerta |

## Introduction.

Here I propose a common neural hub, the MHb-IPN circuit, that explains how diverse sedative, anesthetic, and analgesic agents converge to induce a state of unconsciousness. Opioids and hypnotics inhibit pain but decrease respiration and upper airway muscle activity in anesthetized states (Drummond, 1989; Eastwood et al., 2002, 2005). It is considered unknown why their high doses can stop breathing. Or what causes SWS and anesthesia, besides activated ventrolateral preoptic nucleus (VLPO) and inhibited arousal regions. I introduce new neural mechanisms for how the MHb-IPN system causes loss of awareness, anesthesia, slow-wave sleep (SWS), hippocampo-cortical memory replay, opiates-induced respiratory depression (OIRD), stress-sleep-resilience interactions, rest, recovery, repair, protein synthesis, synaptogenesis, neural growth, dentate gyrus neurogenesis, and protective immune response. Medial habenula has the densest mu-opioid receptors (MORs) in the brain (Gardon and Faget, 2014). I proposed that mu-opioids, glutamate, acetylcholine, nicotine, posterior septum, and bed nucleus of anterior commisure (BAC, amBNST) activate MHb, while the medial septum (MS), horizontal limb of the diagonal band of Broca (hDBB), locus coeruleus (LC, noradrenaline) and ventral tegmental area (VTA, dopamine) inhibit it (Vadovičová, 2015). I described how the MHb promotes rest, recovery, SWS, memory replay, and anesthesia by activating the IPN which activates serotonin release in the median raphe nucleus (MRN). I claimed that the MHb-IPN system promotes a natural slowdown in respiration and heart rate during SWS. This slowdown is linked to decreased oxygen needs because of movement inhibition.

My circuit-based prediction/claim (Vadovičová, 2015) that mu-opioids activate the MHb→IPN pathway was supported by recent Singhal et al. (2025) optogenetic findings. Their influential experimental work proved that MORs have canonical inhibiting effects in both MHb and IPN postsynaptic neurons, but that presynaptic MORs activation strongly increased evoked excitatory glutamate co-release from cholinergic terminals at the MHb-IPN synapses. I suggest that repeated opioid administration might via cannonical inhibitory MORs gradually decrease substance P release from dorsal MHb neurons, causing tolerance to the same dose, and smaller analgesic effects. I predict that high MOR activation (by optogenetics) in mice MHb will cause opioid-induced respiratory depression OIRD, even respiratory arrest. I suggest that optogenetic activation of specific MHb, IPN, or MRN receptors can test my idea that MOR, acetylcholine, nicotine, adenosine, and many anesthetics activate the MHb→IPN→ MRN→ serotonin→claustrum →cortical SWA circuit. These agents, plus ketamine and 5-HT2a cause loss of awareness by inducing SWA in cortical and striatal neurons, and by IPN-MRN inhibition of theta state and arousal promoting regions.

Inhaled Anesthetics (Sevoflurane, Desflurane, and Isoflurane) increase signals to chloride channels (via GABA receptors) and potassium channels while decreasing signals to excitatory pathways (Miller et al., 2024), Intravenous Anesthetics (Propofol and Midazolam) increase the duration of GABA-activated channel opening, leading to increased chloride influx and hyperpolarization of postsynaptic neuronal membranes (Folino et al., 2024), while ketamine inhibits hyperpolarization-activated cyclic nucleotide-gated potassium channel 1 (HCN1) receptors (Rosenbaum et al., 2024). Thus, I suggest that K channels inhibited by ketamine cause the IPN activation, while presynaptic MORs at MHb-IPN synapses, or anesthetics at IPN might also inhibit certain K channels. I claim that gabaergic IPN and MRN projections inhibit (similar to VLPO) arousal promoting nuclei i.e. vPAG+VTA (dopamine), LC (noradrenaline), LH (orexine), TMN (histamine), DRN, the attention and encoding linked hDBB subgroups (and activate its SWA-on subgroups), theta promoting MS/DBB, SUM, NI, LDT/PPT, and arousal/pain linked PBN. Thus my neural model/hypothesis integrates previous hypotheses with the MHb-IPN circuit and its efferents.

Residual anesthetics may cause postoperative confusion, difficulties concentrating during awakenings, memory loss, weaker theta/gamma power and alertness by returning IPN, MRN, and claustrum activation. The IPN likely attenuates the hDBB wake-on (and activates the hDBB SWS-on) subgroups, arousal circuit, nucleus incertus, norarenaline and dopamine. Partial claustrum activation leads to prefrontal SWA, while serotonergic disruptions of MS/vDBB disrupts theta oscillations and hippocampal encoding (of bound-contextual information).

Long-term sleep disturbances after surgery weaken healing, recovery and immune protection, leading to chronic pain and cognitive decline, while ketamine may uniquely preserve some aspects of restorative sleep (Patel and Ownby, 2025). In support of these findings my paper proposes how ketamine and 5-HT2a agents promote repair, protein synthesis, synaptogenesis, anti-inflammation, and recovery roles of SWS, by activating the IPN→ MRN→ serotonin→ BDNF + melatonin pathways. It further claims that they attenuate postsurgery pain by gabaergic IPN projections to LHb, PAG, and parabrachial nucleus.

Figure 1.
The interactions and effectors of the MHb-IPN circuit cause SWS, loss of awareness, hippocampo-cortical memory replay of bound/linked information, anesthesia, and OIRD.
Figure shows a model/mechanisms how activation of the MHb by mu-opioids, and the IPN by ketamine and phencyclidine causes loss of awareness by activating claustrum which induces cortical and some subcortical slow-wave oscillations. Model proposes that anesthetics directly or via MHb activate the IPN-MRN-serotonin-claustrum-cortical SWA circuit, inhibit via the IPN + MRN efferents the theta states, working memory, arousal and REM sleep, and inhibit dopaminergic

nuclei, DRN, and arousal by activating the IPN-LPO-RMTg efferents. The IPN and MRN inhibit awareness also by inhibiting dopamine which promotes working memory (Goldman-Rakic, 1992) and signals values and meanings (Vadovičová and Gasparotti, 2013, 2014), by inhibiting nucleus basalis of Meynert (NBM) involved in gamma oscillations, multiregional synchronisation, and attention, and by inhibiting arousal-promoting regions (LC, PBN, PAG, LH, PH, NI..).
Model claims that MHb and IPN reciprocally inhibit LHb (medial part mostly) thus attenuate anxiety, affective pain, withdrawal, depression, and neuroinflammation. Serotonin release causes synaptogenesis, BDNF-induced neural growth, and dentate gyrus neurogenesis. I predict that serotonin, ketamine, and phencyclidine activate the 5-HT2a receptors in the IPN and claustrum. They cause the claustrum-induced slow-wave activity (SWA) in cortical and some subcortical regions which leads to loss of awareness. I claimed that mu-opioid receptors MOR activate medial habenula and SWS (Vadovičová, 2015). Anteromedial BNST, also called BAC projects to MHb and induces SWS and memory consolidation in safe enough situations. Dentate gyrus induces hippocampo-cortical memory replay (to avoid interference in incoming information) by activating triangular septum which activates ventral MHb. MRN serotonin then alters hippocampal state into replay and inhibits encoding. Anteromedial BNST activates MHb in safe enough situations, promoting SWS, loss of awareness, and recovery states. Dorsal MHb activated by safety and infection interacts with immune response via glia, and activates vMHb. I claim that stress/worries evoked noradrenaline, and reward and novelty evoked dopamine reciprocally inhibit the MHb-IPN output. Thus, they diminish SWS, its homeostatic functions, and disrupt the inhibitory effect of MHb on LHb. This model proposes that the MHb-IPN circuit attenuates the ventilation/respiration response by inhibiting the hypercapnia-sensing and other DRN neurons, and by interacting with its PAG effectors. Serotonergic DRN (which is more active in wakefulness than in SWS, and down in REM sleep) activates the preBötC (Manzke et al., 2003), and lateral PBN which induces ventilation to hypercapnia (Liu et al., 2021) and arousal to harmful sensations (Kaur et al., 2020).
Yellow arrows show main serotonergic MRN projections. BAC, bed nucleus of the anterior commissure; BNST, bed nucleus of stria terminalis; DRN, dorsal raphe nucleus; locus coeruleus, LC; MOR, mu-opioid receptors; MS/DBB, medial septum/diagonal band of Broca; NTS, nucleus tractus solitarius; PAG, periacqueductal gray; PBN, parabrachial nucleus; preBötC, preBötzinger complex; SWA, slow-wave activity (0.1-2Hz); vACC, ventral anterior cingulate cortex (Brodmann area 25, infralimbic cortex in rats); ventral tegmental area (VTA, dopamine). This figure and model was presented in my Research Gate paper, and at FENS Forum 2024 poster.

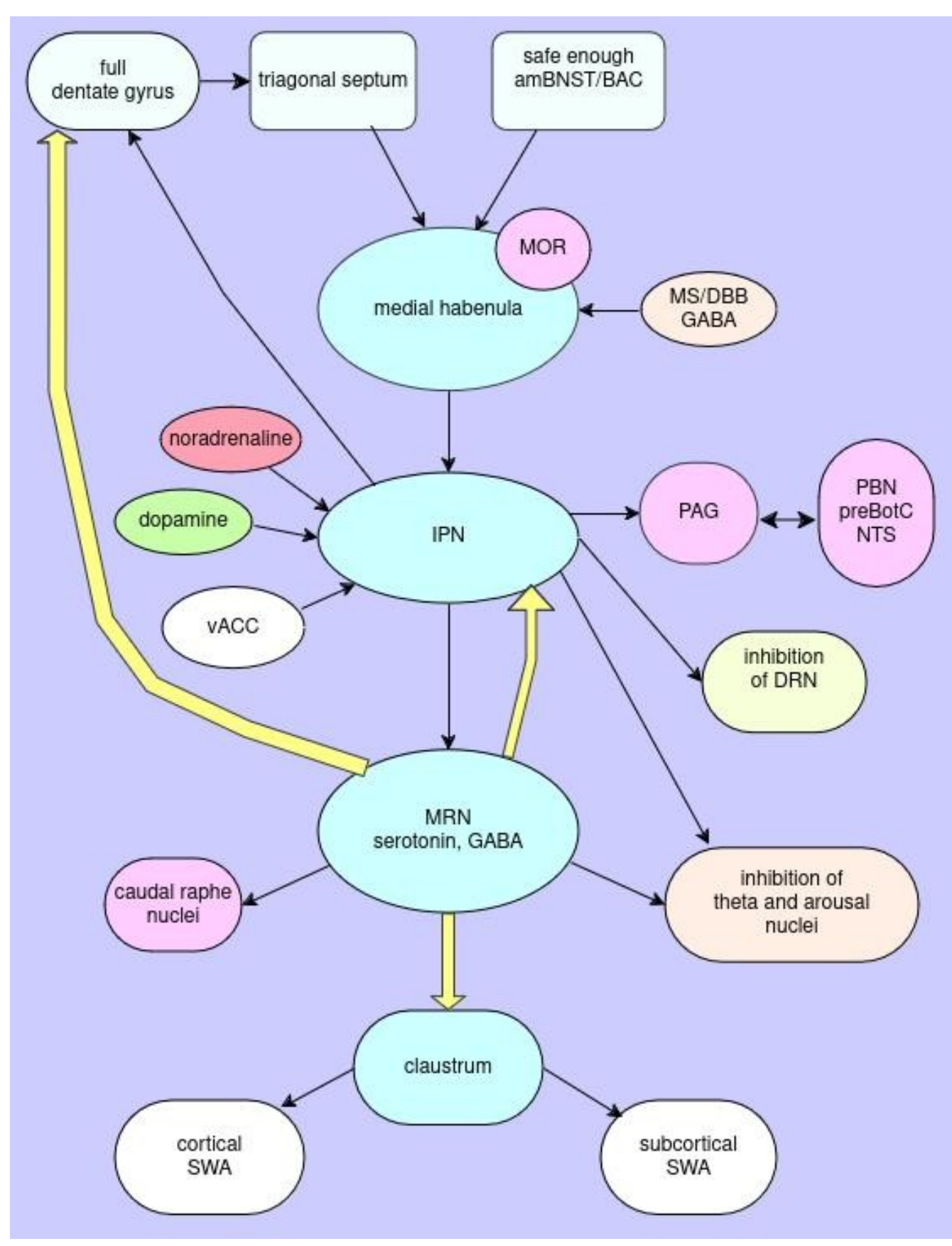

Previously, I proposed that in promoting SWS, the IPN and MRN activate the VLPO which activates the rostromedial tegmental nucleus (RMTg) that attenuates movement by inhibiting dopaminergic neurons: IPN + MRN →VLPO →RMTg →dopamine nuclei down (Vadovičová, 2015). The RMTg inhibits strongly VTA, SNc and DRN (Perroti et al., 2005; Jhou et al., 2009 a, b; Kaufling et al., 2009; Balcita-Pedicino et al., 2011). Cell group a10 contains a dopaminergic VTA and ventral PAG cluster, located between the left and right oculomotor nuclei. So, I propose that RMTg in SWS and anesthesia also inhibits the dopaminergic vPAG and zona incerta (ZI) neurons. Zhao et al. (2022) identified 71 regions projecting to the RMTg, e.g. LHb, LH, ZI, PnO, PAG, superior colliculus, LDT, PPT, IPN, MRN, DRN, VLPO, NAc, CPu, VP, SNr, SNc, VTA, LC, PH, SI, hDBB, vDBB, MMB, l-PBN, motor, somatosensory, cingulate and prelimbic cortex (PFC and dACC homolog), and anterior insula. Thus I predict/propose that the IPN, MRN and VLPO activate RMTg which inhibits dopaminergic vPAG. And that D2 indirect pathway of nucleus accumbens via GPe inhibition disinhibits SNr that activates RMTg linked motor inhibitions, active also in SWS: striatal D2 (receptors) neurons→ GPe→ SNr→ RMTg→ vPAG down→arousal down. My prediction is that noradrenergic LC projections attenuate RMTg output, to bias for active avoidance/response in dangerous situations. Li et al. (2018) discovered that propofol causes anesthesia by increasing the inhibitory GABA release on dopaminergic vPAG neurons, thus inhibiting their activity. Microinjection of a GABA A receptor antagonist disinhibited the dopaminergic vPAG neurons. Liu et al. (2020) found that isoflurane anesthesia causes increased presynaptic inhibitory GABA release in vPAG, and decreases dopamine release from vPAG neurons. The activity of vPAG neurons increased during the emergence of isoflurane anesthesia. The induction of anesthesia was shortened, and the recovery time was lengthened by unilateral ablation of dopaminergic vPAG neurons. Thus I predict that many anesthetics act via GABAergic projections from the IPN to vPAG, and other arousal and theta states promoting brain regions. The IPN induces loss of awareness by activating the MRN→serotonin→claustrum→neocortical slow-waves.

Vetrivelan et al. (2010) proposed that wake-active dopaminergic vPAG neurons promote wakefulness by inhibiting the sleep-promoting neurons in the VLPO, and by stimulating the cortex and the wake-promoting cell groups, including the LH orexin neurons, PPT/LDT and basal forebrain (BF) cholinergic neurons, and PBN/PC glutamatergic neurons (PC, precoeruleus nucleus). Further, they suggested that GPe-cortex loop may relay and enhance cortical arousal, and stabilize excitatory state of pyramidal neurons. By activating the D2 receptors in the striatum, dopamine disinhibits the external globus pallidus (GPe) and promotes cortical activity and behavioral arousal. The cortex-GPe-cortex, cortex-striatum-GPe-cortex, and cortex-striatum-GPe-thalamus-cortex networks link GABAergic GPe with cortex. Their tracing studies found that GPe targets both GABAergic interneurons and glutamatergic pyramidal neurons, mostly in cortical layers IV and V (with reciprocal connections) in rats. Cortical layer V is important for generating EEG waves (Sanchez-Vives and McCormick, 2000). The EEG delta power (linked to NREM/SWS sleep) is increased and theta power (linked to memory encoding, active wake and REM sleep) is reduced after GPe lesions (Qiu et al., 2010). Medial GPe projects to the medial PFC, while lateral GPe targets the insular, motor, and sensory cortices (Saper, 1984).
Cortex→D2R striatum →GPe→ GPi/SNr pathway activates the LHb through GPi input. But I claimed that during SWS is LHb reciprocally inhibited by MHb, IPN and MRN (Vadovičová, 2015). So, I suggest that activation of the striatal D2 indirect pathway (inhibited by dopamine, activated by adenosine) promotes the EEG delta power by activating the SNr which inhibits the VTA dopamine release, and via activating RMTg inhibits noradrenergic locus coeruleus, serootnergic DRN, and dopaminergic nuclei including vPAG, leading to suppression of arousal and locomotion. I suggest that serotonergic deficit in Parkinson's disease diminishes SWS, by overstimulating the LHb. LHb inhibits dopamine, serotonin and noradrenaline release. So, the overactive LHb can cause frequent night-time arousal, day-time sleepiness, and less SWS linked memory replay by inhibiting the IPN and MRN, and disinhibiting D2 indirect pathway that inhibits VTA and activates SNr and GPi→LHb. Volkow et al. (2012) documented that striatal D2 receptors downregulation after one night sleep deprivation in humans was linked to sleepiness. In addition I claim that DRN serotonin inhibits (possibly via 5-HT1B receptors) the D2 indirect pathway in the ventral striatum, thus attenuates (releases from) learned helplessness, social and affective inhibitions, and aversions, and weakens the inhibitory avoidance caused by bipolar disorder down state or depression. Serotonergic DRN arouses cortex and cognition by disinhibiting the SNc release of dopamine to cortex.

Mahon and colleagues performed intracellular recordings from striatal median spiny neurons and showed their rhythmic action potential fluctuations between a DOWN state (hyperpolarized quiescent state) and an UP state (depolarized state) during SWS, coincident with cortical activity. During wake and REM sleep they switched into high rates of discharge with "random" patterns that were not synchronized with cortical EEG (Detari et al., 1987; Mahon et al., 2003; Mahon et al., 2006). Also GPe neurons are more active during wake and REM sleep (Urbain et al., 2000). Gerashchenko et al. (2008) identified, in mice, rats, and hamsters, a population of cortical and some caudate-putamen GABAergic projection interneurons, active in SWS in proportion to homeostatic sleep drive, and expressing the enzyme neuronal nitric oxide synthase (nNOS). The Fos expression (activity marker) in this nNOS expressing population with cholinergic and serotonergic afferents was highly correlated with the slow-wave activity (SWA) intensity and slow-wave sleep amount, so they proposed that this nNOS neurons subset is involved in SWA and homeostatic sleep regulation. Further, they found (Dittrich et al., 2012) that these cortical nNOS neurons co-express the Substance P (NK1) receptor and are depolarized by Substance P. Acetylcholine from basal forebrain cells largely excited cortical nNOS/NK1R cells, while reducing glutamatergic input onto these neurons (Williams et al., 2017). Somogyi et al. (2012), found that most NO interneurons with soma in the stratum radiatum of the CA1 area exhibit characteristic properties of ivy cells and express neuropeptid Y. Using the phototoxicity of a NO indicator, selective damaging of cortical NO interneurons uncovered their role in lateral inhibition on neighboring columns and in the spatiotemporal dynamics of cortical activity (Shlosberg et al., 2012).
These SWA promoting projection interneurons enable memory replay state, in which the hippocampal replay of encoded memories - sequences of neuronal assemblies (co-activations) bound by hippocampus re-activate cortical neurons and synapses via cortical ripples (brief 120 - 200 Hz oscillations) observed mostly at the UP states of cortical and subcortical SWA. Perhaps the day-time accumulated adenosine, known to promote sleep, might activate the Substance P releasing medial habenula, leading multisynaptically to activation of cortical and (some) subcortical nNOS/NK1R cells. The MHb and IPN receive GABAergic projections from hDBB and medial septum (Agostinelli et al., 2020). The IPN might interact with the SWA-active nNOS/NK1R interneurons via hDBB. The source of Substance P and serotonin at nNOS/NK1R interneurons is unknown.
I propose that during SWS the MRN serotonin, glutamate and/or gabaergic terminals inhibit some subset of cortical or hDBB interneurons that activate gamma oscillations (of 20-100 Hz), but activate those interneuron populaitons that support SWA and cortical ripples (high frequency oscillations 120 - 200 Hz) which potentiate the re-activated synapses. My model predicts that higher serotonin and ketamine levels activate the IPN which activates MRN. IPN or MRN might via hDBB (MRN also directly in cortex) synchronize UP and down states (SWA) in the retrosplenial cortex. There are direct IPN projections to ventral dentate gyrus and hDBB, so they might initiate SWR and influence the cortical ripples during SWA. Again, I propose that IPN activates that subset of hDBB projection neurons that promotes hippocampo-cortical memory replay and ripples, but that IPN suppresses those (likely parvalbumine) interneurons populations that support the theta-gamma coupling (5-9 Hz and 20-100 Hz) and memory encoding state. The IPN→ MRN→ claustrum→ cortical/subcortical SWA circuit activates specific populations of cortical interneurons. Jackson et al. (2018) found that in certain cortical areas, claustrum neurons primarily project to parvalbumin and neuropeptide Y GABAergic interneurons.

The somatostatin expressing GABAergic neurons in basal forebrain (BF) were found to inhibit BF cholinergic, parvalbumine, and glutamatergic neurons via GABAergic synapses (Xu et al., 2015). In addition, many of these cells were sleep-active and sleep promoting, and might suppress all wake-promoting BF cell types during non-REM sleep. SWA was found to facilitate synaptic consolidation or produce synaptic downscaling, thereby increasing signal-to-noise ratios in relevant neural circuits (Tononi et al., 2006). There is stronger DRN than MRN input to prefrontal and entorhinal cortex, so there need to be a way to disentagle their opposite effects on cortical activity in wakefulness and SWS. Suppression of the MS/DBB and theta state by the MHb-IPN-MRN pathway inhibits the memory encoding state (in which the hippocampus bounds spatial, temporal and relational information together) to separate it from the memory replay state (SWA, ripple and cortical spindles activity) to not mix unrelated events and information.

The MHb-IPN circuit involved in SWS and anesthesia might activate the parasympathetic system, causing the natural slowdown in respiration and heart rate, and the opioids evoked respiratory depression (OIRD). In addition, the multisynaptic MHb- and SWS-linked activation of RMTg leads to the suppression of DRN. Serotonergic DRN has chemoreceptors sensing hypercapnia. Chemogenetic activation of the lateral parabrachial nucleus (PBL), or its 5-HT2a receptors was proved to increase ventilation, while its deactivation led to OIRD (Liu et al., 2021). I propose that inhibition of DRN by RMTg during the SWS and anesthesia weakens the DRN response to hypercapnia. I claim that MHb via IPN attenuates/inhibits DRN and its chemosensitive response to increased carbon dioxide in the blood - acidity. This decrease in serotonergic DRN output causes lower PBL activation, leading to less hypercapnic ventilation in SWS and anesthesia. DRN serotonin might also activate (in wakefulness) or fail to activate (being inhibited by IPN in SWS) the preBötzC. On the contrary, the MHb-IPN-PAG system likely attenuates/inhibits the PBL and preBötzC activation, causing the OIRD in SWS and anesthesia. The GABA A receptors are involved in anesthesia and I propose that GABAergic RMTg projections inhibit DRN and a10 in vPAG: MHb-IPN-MRN-VLPO-RMTg→ a10 in vPAG + DRN inhibition.

During SWS, parasympathetic efferents are activated more than the sympathetic efferents. Sympathetic output increases alertness, alarm, attention, speed, and strength. The SWS/NREM sleep is linked to safety, rest, regeneration, memory replay, and immune response. Slow-wave sleep is disrupted in cases of danger, threat, harmful, or strong sensory stimuli. The stress, danger, adversities, noradrenaline, and pain diminish SWS, to prod us (humans and mammals) to escape, fight or flight to avoid loss or damage. Previously, I proposed how mu opioids activate MHb, and via the MHb→IPN→MRN→serotonin circuit induce the SWS and unawareness, and how they inhibit the theta states, alert wakefulness, and REM sleep circuits. Thus, the mu-opioids stimulate the MHb→IPN→MRN circuit. This causes strong suppression of alert wakefulness and suppression of those circuits, that are linked to alarm, action, movement, fight, flight, and higher oxygen and respiration requirements: for example the PAG, LC, lateral and posterior hypothalamus (LH, orexin and PH, histamine), pedunculopontine tegmentum (PPT), PVH, and central nucleus of amygdala (CeA.)

I propose a new circuit for respiratory slowdown linked to SWS, the MHb→IPN→PAG→NTS circuit. Over-activation of this circuit by high mu-opioids can cause respiratory depression and arrest. I propose that the mu-opioids activate MHb which activates IPN, which causes respiratory slowdown via inhibiting DRN and/or affecting PAG. IPN and MRN activation inhibits arousal and awareness in SWS and anesthesia,  decreasing the speed of waking up after hypercapnia. Likely, the IPN attenuates the PAG nuclei linked to increased respiration, pain, arousal, and alarm reactivity, and activates those that induce respiratory slowdown. The dlPAG inhibition would decrease the alarm functions, and support the respiratory slowdown. The PAG  receives afferents from nociceptive neurons in the spinal cord and PBN, and projects to thalamic nuclei, that relay nociceptive signals to the posterior insula. The dlPAG activation by pain,  harm, threat (input from amygdala), and sudden sensory stimuli that can signify acute danger leads to alarm, body mobilization for fight or flight, faster respiration to cover the higher needs, and increased alertness. The IPN might also act via other than dlPAG regions. The ventrolateral PAG is linked to fear-induced freezing and temporary inhibition of respiration. Part of PAG works as a descending pain inhibitory system. Its activation inhibits nociceptive neurons in the dorsal horn of the spinal cord (Behbehani, 1995). So mu opioids in MHb likely indirectly activate this descending pain inhibitory system. The dorsal column mediates pressor and the ventrolateral column mediates depressor responses (Behbehani, 1995).

During SWS, strong sensory stimuli, too much carbon dioxide, or insufficient oxygen in blood, increase respiration and heart rate via sympathetic and other regulatory systems. The unnaturally high opioid dose might suppress these alarm/alert reactions of dlPAG, by activating MHb→IPN pathway. The IPN might inhibit the dopaminergic vPAG (involved in increasing arousal to pain or hypercapnia) or modulate other ventral PAG nuclei involved in analgesia and respiratory depression. The IPN→PAG pathway supporting respiratory slowdown might involve the nucleus tractus solitarius (NTS). NTS has afferent and efferent projections to and from all respiratory motor nuclei, is connected to majority of modulatory respiratory regions, and receives visceral afferents and chemosensory input.
This MHb→IPN circuit-based model is highly applicable and testable by optogenetic, neuroimaging, and neuropharmacological methods. In addition, using the DTI method, I found axonal projections between the MHb and pineal gland in humans (Vadovičová, 2014), that might via µ-opioid receptors on MHb activate melatonin production in the pineal gland. Melatonin is known, for moderate inhibition of respiration. The DTI method does not specify a direction of axonal projections, but MHb projects to and activates the pineal population of silent cells (Axelrod, 1970; Ronnekleiv and Moller, 1979, Ronnekleiv et al., 1980), known to produce melatonin in the dark period. So I propose that µ-opioids and other anesthetic agents activate MHb-IPN system and cause loss of awareness, SWS, anesthesia, and respiratory slowdown/depression (OIRD) by acting via the following effectors:

MHb →IPN → MRN→ serotonin→ claustrum → cortical/subcortical slow-waves (SWA).
Loss of awareness in slow-wave sleep and anesthesia.

MHb →IPN →MRN→ they inhibit awareness, arousal, working memory, hippocampal encoding, theta + gamma coupling. To enable hippocampo-cortical memory replay and restorative SWS functions. Down MS/DBB, SUM, MMB, LTD, PPT, SI, NBM, LH, PH, LS, LC, VTA, ZI, NI, LHb.

MHb →IPN →PAG → down lateral PBN + preBötC + NTS.
Respiratory depression by attenuation of the respiration stimulating nuclei.

MHb → IPN → DRN down → lower ventilatory + arousal response to hypercapnia by lateral PBN.

MHb → IPN → vPAG → inhibition of arousal + wakefulness regions: NBM, SI, PPT, LDT, LH, PBN.

MHb → IPN → VLPO up → RMTg up → suppression of dopaminergic nuclei, DRN, and LC.

MHb up → IPN up → RMTg up → down vPAG + VTA + SNc + DRN + LC. Down arousal.

MRN → VLPO up → RMTg up → vPAG down → down NBM, SI, PPT, LDT, LH, PBN + arousal. RMTg inhibits via GABA A receptors the dopaminergic group a10 in vPAG. Also VLPO inhibits vPAG. vPAG inhibition strengthens anesthesia by weakening wakefulness promoting regions.

MHb → IPN → MRN → caudal raphe nuclei + ventromedial PAG → which roles ? analgesia?

I claim that inhibition of dopaminergic vPAG neurons by GABAergic RMTg (via GABA A receptors) inhibits arousal. Dopaminergic vPAG is known to activate wakefulness by activating the NBM (gamma synchronization, working memory, cortical coupling), cholinergic PPT/LDT, orexinergic LH, cortex, and PBN, while inhibiting SWS-promoting VLPO. Photostimulation of basal forebrain GABAergic terminals in the thalamic reticular nucleus (TRN) strongly promoted cortical activation and behavioral emergence from isoflurane anesthesia (Cai et al., 2023). Interestingly, mu-opioids activate RMTg via MHb-IPN circuit in SWS and anesthesia but attenuate spontaneous RMTg activity and GABAergic release from RMTg in VTA/SNc during wakefulness, when they disinhibit dopaminergic nuclei (and serotonergic DRN). Morphine induced a 46% inhibition of IPSCs evoked from the RMTg in dopaminergic VTA in the Matsui et al. (2014) study, and this inhibition of GABAergic input from RMTg showed tolerance to repeated morphine.

Rationale and Aims of this IPN-MHb model.

Pharmacological inhibition of the MRN autoreceptors with serotonin1A agonists, 8-OH-DPAT or buspirone, produces hippocampal theta rhythm in the rat (Vertes et al., 1994). Inhibition of the MS/vDBB (medial septum/vertical limb of the diagonal band of Broca) through serotonin 5-HT2C receptors causes dose dependent suppression of hippocampal theta oscillations (Hajos, 2003). Theta power (4-9 Hz) is strongest in alert wakefulness and REM sleep. While theta states inhibition by MRN, and the ascending arousal system inhibition by VLPO provide known framework for sleep-wake regulation, the specific neural regions that initiate the slow-wave sleep state and, critically, are co-opted by general anesthetics to induce unconsciousness, remain incompletely defined. Furthermore, the lethal link between opioid-induced analgesia and respiratory depression (OIRD) suggests a shared neural substrate that has yet to be fully elucidated. This paper proposes a unifying hypothesis that positions the MHb and its primary effector, the IPN, as a critical node governing state transitions into SWS and anesthesia. This work aims to combine anatomical and functional evidence to demonstrate that the MHb-IPN circuit:

1. Acts as a hub for inducing cortical slow-wave activity (SWA) and loss of awareness via the IPN → MRN → claustrum projections to cortical and striatal neurons.

2. Provides a novel circuit-based explanation for OIRD through its inhibitory influence on respiratory (IPN → PAG → preBotC) and chemosensitive (DRN) brainstem nuclei, and arousal.

3. Offers a neural pathway for the convergent or parallel actions of diverse agents that activate - mu-opioids, cannabinoids, acetylcholine, nicotine, adenosine, substance P, ketamine, GLP-1RA, psychedelics, anesthetics, or inhibit it – noradrenaline, dopamine, stress, worries, novelty, danger.

By integrating information from anatomical, neuropharmacological, and functional studies across few species, this model aims to generate a novel, testable framework for understanding the neural basis of loss of awareness, and its associated physiological changes. The proposed mechanisms and predicted neural interactions are inferences based on synthesis of known anatomical and functional findings. To ensure clarity and testability, the model's core predictions are explicitly stated throughout the text and summarized in the conclusion. This SWS model does not preclude the existence of parallel or redundant mechanisms. Its primary value lies in its ability to generate falsifiable hypotheses for future experimental investigation.

## Theoretical Framework: Methodological Framework.

This modeling approach followed systematic principles:
1. Circuit Reconstruction: Monosynaptic projections were verified by multiple tracing studies
2. Functional Validation: Only physiologically demonstrated effects were incorporated
3. Pharmacological Plausibility: Receptor distributions matched proposed drug actions
4. Evolutionary Conservation: Included mechanisms were conserved across rodents and primates
5. Clinical Consistency: Predictions aligned with established clinical observations

When contradictory evidence existed, the best-replicated findings were adopted, and discrepancies explicitly noted as limitations.

## Predicted Neurochemical Mechanisms and Receptor Specificity.

The receptor-specific topography enables distinct pharmacological profiles while maintaining circuit-level integration. The MHb-IPN-MRN circuit exhibits remarkable neurochemical heterogeneity that underpins its proposed functional diversity. To clarify the specific mechanisms, my model proposes the following receptor-level interactions:

• MHb: Substance P (SP) release from dorsal MHb subregion, activated by amBNST/BAC input, might signal that it is safe enough to fall asleep. Repeated mu-opioid dose might postsynaptically decrease the SP release, and cause tolerance. Question is if the SP pathway from MHb influences different regions than the acetylcholinergic, or the IL-18 releasing pathway. Neuropharmacology might target them specifically. Question is from where comes the SP that interacts with cortical nNOS/NK1R interneurons, active in SWA.

• MHb: Some of the calming and anti-inflammatory effects of endocannabinoids are caused by its disinhibition of MHb, leading to sleep- and serotonin-linked repair/recovery. Cannabis-1 receptor (CB1-R) activation suppresses theta and gamma activities in awake rats during exploratory behavior (Hajos et al., 2008; Robbe et al., 2006), and induces trains of high-voltage spindles during immobility (Sales-Carbonell et al., 2013). I propose that cannabinoids cause this suppression by activating the MHb-IPN-MRN circuit. The IPN→ hDBB→ TRN (thalamic reticular nucleus) projections then synchronize spindle generation. Possibly via the SWA-active subgroup of hDBB neurons. Agostinelli et al. (2019) found acetylcholinergic and gabaergic hDBB projections to TRN.

• IPN: I predict that ketamine, phencyclidine, anesthetic 5-HT2a agents, and nitrous oxide drive MRN activation (and its serotonergic + gabaergic terminals), via 5-HT2a receptors (or their signalling cascade) in the lateral IPN. This activated IPN then activates MRN. Ketamine and nitrous oxide might act via signaling cascade of 5-HT2a receptors, such as G-proteins linked K channel inhibitors.

• MRN: MHb, IPN and MRN activation has antidepressant effects, as each mutually inhibits the LHb (or at least its MRN inhibiting medial part). This enforces the next day resilience, effort, locomotion, well-being, via disinhibition of dopamine and serotonin release. MRN serotonin decreases neuroinflammations (during SWS), while MHb and IPN inhibit LC noradrenaline, VTA dopamine, and stress. Thus activation of MHb decreases negative affects of withdrawal.
• Claustrum: The IPN activates MRN. Serotonergic MRN projections activate via 5-HT2a receptors claustrum. (Claustrum was found to induce cortical slow-wave activity through claustral projections to various groups of GABAergic and pyramidal neurons.)
• Cortex: Hypothesis that Substance P or serotonin activate the cortical or striatal nNOS/NK1R interneurons possibly involved in ripples (100-200 Hz), and known to be active only during SWA.

The model incorporates critical dose-dependent dimensions:
• Acute opioid exposure (minutes): Presynaptic MOR facilitation dominates, enhancing MHb-IPN transmission. Sleep initiation is likely dose dependent
• Chronic opioid exposure (days): Postsynaptic 'cannonical' inhibition via MOR might decrease the MHb-IPN activation and lead to tolerance. It might also lower substance P input to MHb
• Low-dose ketamine (0.5-1 mg/kg): Predominant vACC→DRN activation (via 5-HT2a receptors on vACC/a25) promotes antidepressant effects.
• Anesthetic ketamine (5-10 mg/kg): IPN→MRN→claustrum activation induces cortical SWA and loss of awareness, while IPN inhibits DRN.
These dynamics explain why different behavioral states emerge after ketamine or 5-HT2a administration under varying pharmacological conditions. Model proposes the dose-dependent effects of ketamine. Low dose promotes cognition: vACC→ DRN→ SNc→ PFC; while high dose induces anesthesia: IPN→MRN→ claustrum→ cortex.

## The anatomical and functional evidence for the proposed mechanisms:<br>Medial habenula connectivity and functions.

Herkenham (1981) found spared or increased metabolism (evidenced by [14C] deoxyglucose intake) in the MHb, IPN, and the habenulo-interpeduncular tract after applying anesthetics in rats. The pentobarbital, ether and chloral hydrate reduced deoxyglucose uptake in most neuronal regions. So anesthesia might activate the same MHb→IPN circuit which induces SWS and respiratory slowdown as do mu-opioids.
The MHb has afferents from the glutamatergic and ATP-ergic triangular septum, cholinergic and ATPergic septofimbrial nuclei, GABAergic medial septum and vertical diagonal band of Broca (MS/vDBB), noradrenergic locus coeruleus (LC) and sympathetic superior cervical ganglion, substance P-ergic anteromedial part of BNST (amBNST or BAC, bed nucleus of anterior commissure), dopaminergic ventral tegmental area (VTA) and median raphe nucleus (MRN), (Sperlágh et al., 1995; Herkenham and Nauta, 1977; Gottesfeld, 1983; Qin and Luo, 2009; Lecourtier and Kelly, 2007; Phillipson and Pycock, 1983; Sutherland, 1982; Conrad et al., 1974; Yamaguchi et al., 2013).
MHb is known for dense mu-opioid receptors (MOR). Gardon et al. (2014) generated a MOR-cherry knock-in mouse line. They found that MOR's expression in the basolateral MHb was co-localized with a subpopulation of cholinergic neurons, and with most of the substance P neurons in the apical MHb. The MORs were also present in the substance P and cholinergic terminals in the central and intermediate IPN, and in the septal regions projecting to the MHb.
Chunxiao et al. (2023) found that the Neuropeptide Y (NPY) produces analgesic and anxiolytic effects through the Y1 receptors in the MHb, a possible therapeutic target for the treatment of migraine. I add that focused natural or artificial light exposure on retina might feel noxious also through evoking sympathetic system noradrenaline release which inhibits the MHb output. This inhibition of MHb then indirectly decreases MRN serotonin release in melatonin producing pineal gland. Noradrenaline also hinders sleep by inhibiting melatonin. I suggest that NPY and other MHb-activating agents decrease pain through inhibitory effects of MHb, IPN and MRN on LHb, and through MHb→ IPN→ analgesic pathway.
I suggest that the MHb-IPN projections calm the pain pathway (possibly via PAG→PBN) to enable sleep and recovery. Loss of awareness induced by the MHb-IPN system might also be a way out from pain, via endorphins. Also substance P, released by superior MHb, might decrease pain, as substance P receptors were found in spinal dorsal horn, NTS, PBN, and PAG which all react to pain. I claim that medial habenula calms anxiety, depression, affective pain, and withdrawal by inhibiting the LHb directly, and via IPN and MRN.

Aizawa et al. (2012) showed that different subregions of the MHb use different neurotransmitters and could be categorized as either exclusively glutamatergic (superior part of MHb), both substance P-ergic and glutamatergic (dorsal region of central part of MHb), or both cholinergic and glutamatergic (inferior part, ventral region of central part, and lateral part of MHb). He found that the superior part of the MHb strongly expressed interleukin-18 and was innervated by noradrenergic fibers. So noradrenaline induced by chronic stress and worries can attenuate the immune defense reactions to microbes linked to MHb-IPN-MRN and SWS. Thus I claim that chronic stress increases detrimental inflammations and oxidative stress also by lowering the MHb, serotonin→melatonin, and serotonin→ BDNF→ proteins synthesis linked restorative, regenerative, anti-inflammatory and immune functions of SWS. MHb sends topographically organized glutamate, substance P, and acetylcholine output to IPN through the fasciculus retroflexus (Qin and Luo, 2009; Ren et al., 2011; Herkenham and Nauta, 1979). MHb also projects to the pineal body (Ronnekleiv and Moller, 1979), LHb, and sparsely to the VTA (Cuello et al., 1978), RMTg, DRN, and MRN (Sutherland, 1982). I suggest that melatonin stimulation by MHb and MRN serotonin has protective effects against noxious neuroinflammations and some neurodegenerations: MHb→ pineal gland→ melatonin up→ protective immunity.

I propose that endocannabinoids attenuate the theta states (also hippocampal memory encoding), arousal, and stress, and promote relaxation, SWS, and LHb inhibition by disinhibiting the MHb-IPN-MRN circuit. Vickstrom et al. (2021) showed that endocannabinoids decrease the GABAergic inhibition of MHb by MS/DBB. With RNAscope in situ hybridization, they found that key enzymes that synthesize or degrade the endocannabinoids 2-arachidonylglycerol (2-AG) or anandamide are expressed in the MHb and MS/DBB, and that cannabinoid receptor 1 (CB1) is expressed in the MS/DBB. Their electrophysiological recordings in MHb neurons revealed that endogenously released 2-AG retrogradely depressed GABA release from the MS/DBB. Thus I claim that mu-opioids, nicotine, endocannabinoids, and anesthetics activate the MHb-IPN system (via MHb or IPN) which through its efferent projections causes loss of awareness. Activation of the MHb-IPN-MRN-claustrum pathway induces cortical and subcortical slow-wave activity. Activation of the MHb, IPN and MRN inhibits the theta states, arousal, working memory, new memory encoding, alert wakefulness, awareness, and the REM sleep promoting regions (dopamine burst firing, noradrenaline, MS/DBB, SUM, NI, MMB, LDT, PPT, NBM, LH, PH, LC, VTA, and more). Anesthetics also act by activating the IPN and its effectors.

## Interpeduncular nucleus: connectivity and functions.

Shibata et al. (1986) found the afferent projections to the IPN from the nucleus incertus (NI), dorsal tegmental nucleus of Gudden (DTg), LDT, ventromedial PAG, MHb, DRN, MRN, supramamillary nucleus (SUM), preoptic nucleus and DBB. IPN afferents were also confirmed from the MS/vDBB and horizontal DBB (Contestabile and Flumerfelt, 1981), substantia innominata (Vertes and Fass (1988), infralimbic cortex (Takagishi and Chiba, 1991), hypothalamus (LH, PH?) and SUM (Contestabile and Flumerfelt, 1981; Hamill and Jacobowitz, 1984), raphe nuclei (Conrad et al., 1984), NI and DTg (Hamill and Jacobowitz, 1984). Dopaminergic afferents to IPN, mostly from VTA and the medial tier of the SNc, were found to activate the IPN neurons with D1 receptors but to attenuate the majority of IPN (Szlaga, 2022). MHb projection to the paramedian IPN and deep tegmental nuclei, and LHb projections to the apical and central IPN, and DTg were found in cats in Smaha and Kaelber (1973) studies.
So the question is what are the vACC/infralimbic cortex (vACC/a25/IL) projections to the IPN doing? Ventral ACC signals safety and well-being (when reaching the 'comfort zone' of homeostasis and survival (Vadovičová and Gasparotti, 2014). vACC projects directly to the PAG and lateral parabrachial nucleus (Takagishi and Chiba, 1991) which are part of the pain pathway. I predict/claim that the vACC/IL activates the IPN→MRN→hippocampus + claustrum pathway, causing the cortical slow-wave oscillations (SWA), hippocampal sharp wave ripples, and hippocampo-cortical memory replay + cortical ripples during quiet wakefulness, both in animals and humans. These brief cortical SWAs would cause moments of awareness lapses in humans. Mostly in those intervals not occupied by thoughts, or by recording new information, thus during boring or repetitive information moments. I also predict that vACC/a25 via its PAG and lateral PBN projections activates the analgesic pathway which decreases the incoming pain signal (from the body). I propose that vACC/IL attenuates the recording of new information and events linked to theta state, by directly inhibiting SUM and MMB (medial mamillary bodies). I propose that vACC promotes safety, rest, and consumatory behavior linked hippocampo-cortical memory replay in rats by activating IPN. I predicted that vACC/IL calms the affective pain and bad, unpleasant feelings (sick, disgusted) by inhibiting the anterior insula, and calms stress, worries, and threats by inhibiting dACC/prelimbic, CeA, LC, and PVN, and activating the medial or inhibiting the lateral BNST. I think that after a big meal can vACC/IL induce a nap, by activating the LPO which then activates RMTg and inhibits dopaminergic nuclei. The cited infralimbic cortex efferents were found by Takagishi and Chiba (1991) and strong infralimbic cortex to SUM fibres also by Hayakawa et al. (1993).
I have an explanation for Szlaga (2022) findings about dopaminergic afferents to IPN: If recorded D1 receptors were located on the GABA-ergic IPI subregion of IPN, dopamine would activate the inhibitory IPI neurons but suppress the main IPN output.

A recent study with retrograde and anterograde tracers confirmed reciprocal projections between the LDT and IPN, MRN, NI, SUM, and LHb (Bueno et al., 2019). Because the LDT promotes arousal, wakefulness, and theta states, I propose that the LDT reciprocally inhibits the IPN and MRN, and activates the theta promoting NI and SUM. The same group showed that input to IPN was strong to moderate from the prelimbic (human dorsal ACC), infralimbic (human ventral ACC), and cingulate cortex (homolog of the human prefrontal cortex), MHb, MRN, DRN (mostly caudal), LDT, NI, LC, hDBB, SUM, septofimbrial nucleus, LH/PH, dlPAG, lPAG, vlPAG and spheroid nucleus, weak from the paramedian raphe nucleus, MPO, LPO, sparse from the raphe interpositus, mesencephalic reticular formation, retrorubral field, supralemniscal region with serotonergic B9 cell group, dorsal nucleus subcoeruleus, septohippocampal nucleus, ventral LS, pedunculopontine nucleus PPT, paraventricular thalamic and hypothalamic nuclei (Lima et al., 2017). They found that known bidirectional connections between the IPN and MRN are primarily GABA-ergic.
They showed ascending dorsocaudal IPN projections to the ventral hippocampus (ventral CA1 field and, more caudally, the molecular and granular layer of the dentate gyrus), caudal ventral LS, SUM, hDBB, and LHb, strong descending projections to dlPAG, lPAG, vlPAG, MRN, lateral and caudal DRN, NI, and LDT, and sparse to interfascicular nucleus, paratenial thalamic nucleus, MPO and the rostral linear raphe nucleus.
I predicted that the IPN mutually inhibits the LHb and the theta, arousal, wakefulness, and REM activating regions: NI, SUM, LS, hDBB, nuclei of Gudden, LH, VTA, and LDT, while activating the slow-wave sleep, MRN and hippocampal sharp wave ripples (Vadovičová, 2015). So by suppressing the theta, alert wakefulness, REM sleep, and arousal promoting regions/circuits, and by activating the MRN-claustrum pathway the IPN enables slow-wave state linked to uninterrupted memory replay, loss of awareness and SWS. I propose that IPN also inhibits the dopaminergic vPAG neurons, midline centromedian thalamus, and parafascicularis thalamic nuclei, and that MRN's serotonin induces synaptogenesis, neural growth, dentate gyrus neurogenesis. Synaptogenesis strengthens the newly encoded temporal, spatial and associative links between information in event-based and relational memories.

Groenewegen et al., (1986) found cell bodies immunoreactive for substance P in rostral IPN, for enkephalin in rostral, apical, and caudodorsal central IPN, and serotonin-containing neurons in the apical subnuclei and the caudal part of lateral IPN. They found main IPN efferents to MRN, DRN, caudal part of PAG, ventral and dorsal tegmental nuclei of Gudden (VTg and DTg), minor projections to the mediodorsal nucleus, the nucleus gelatinosus, midline centromedian and parafascicular thalamic nuclei. The IPN efferents also project to the preoptic area, hippocampus and entorhinal cortex, LH, PH, VTA, DRN, MRN, midbrain raphe, LS, DBB, and LHb (Shibata and Suzuki, 1984; Massopust and Thompson, 1962; Morley, 1986; Smaha and Kaelber, 1973). Medial LHb projects to MRN (Behzadi et al., 1990) and MRN to dorsal hippocampus (Vertes et al., 1999).

Special attention goes to the IPN projections to hippocampus and entorhinal cortex, as the fast-working ATP- and glutamatergic terminals from the triagonal septum (which has dentate gyrus input) and the ATP- and acetylcholinergic terminals from the septofimbrial septum likely induce via MHb-IPN pathway the hippocampal sharp wave ripples (memory sequences of information bound together by interrelations, time and space) that activate hippocampo-cortical replay through cortical ripples, thus consolidation of memories bound by CA1 and CA3 recurrent networks. IPN projects to both the ventral hippocampus and the MRN-claustrum-cortex pathway. So IPN can promote cortical slow-wave oscillations (SWA) and set up the 'replay state' that enables cortical ripples during the up-states of SWA. Baisden et al. and Wyss et al. (1979) demonstrated the IPN and MRN projections to the hippocampus.

Lima et al. (2017) found dense MRN axons in IPC and IPI subregions of IPN that themselves have few terminals in MRN. They found that the IPDM and IPL subregions which have robust feedforward projections to MRN, have scarce MRN feedback.
Thus I predict that the IPI subregion is mostly GABAergic and that GABA-ergic MRN feedback to IPI disinhibits the main IPN output. This enforces the IPN signal which inhibits the theta-state regions and activates the SWS-inducing regions. In line with my claim are Lima et al. (2019) findings that tracer in the IPL part of IPN led to robust septal, hippocampal, and hypothalamic labeling similar to that in IPR. All these ascending projections were meager or undetectable with a tracer confined in IPI/IPC subregions. Other supporting evidence comes from my interpretation of the Szlaga (2022) findings. She demonstrated that optogenetic stimulation of dopaminergic VTA terminals in the IPN led to D1 receptor-dependent increase in caudal IPN activity and c-Fos immunoreactivity specifically in D1R-expressing IPN neurons even though it caused a decrease in the number of IPN c-Fos+ neurons when compared to the controls.
So if VTA dopamine axons activate mostly the GABAergic IPI subregion, it would cause IPI activation but deactivation of the remaining IPN subregions, thus lowering the main IPN output. This IPI activation by dopamine leading to IPN suppression would decrease the cortical slow-wave oscillations and increase the theta-state linked arousal, wakefulness, new memories encoding, and REM sleep. So the MRN feedback to the IPI might do the opposite - inhibit the IPI, thus disinhibit the IPN. I propose that dopamine activates the GABA-ergic neurons in the IPI subregion which consequently attenuates the main IPN output, as both a novelty, new meanings and a chance/expectation to gain rewards make us excited and unable to fall asleep. Also, danger/worries keep us alert/awake, and via the dACC→LC→noradrenaline they inhibit the MHb and IPN, hindering cortical SWA and SWS sleep.
Dopamine signals novelty, value, and meanings, enforces working memory representations, fuels motivation and locomotion, activates theta-circuit (i.e. SUM) and arousal, it is involved in REM sleep functions. Thus dopamine is important in awareness and conscious processing.

## Median raphe efferents and the MHb→IPN→MRN functions.

Electrical stimulation of the MRN directly blocks the hippocampal theta oscillations (Vertes, 1981), and injections of pharmacologic agents that inhibit the activity of serotonergic MRN neurons (into MRN) generate theta rhythm (Vertes et al., 1994; Kinney et al., 1994, 1995) at very short latencies 30–120 seconds and for very long durations 45–90 minutes in rats. Similar findings have been described in the behaving cat (Marrosu et al., 1996). Vertes et al. (1999) found MRN efferents to the caudal raphe nuclei, lateral DRN (less in medial), ventromedial PAG, DTg, IPN, medial mammillary body (MMB), SUM, MS/vDBB, hDBB, hippocampus (to dorsal and ventral HF, DG, CA3 and CA1, and much more than from DRN), claustrum, LDT, PPT, PH, LH, vLS (strong), iLS, dLS, BNST, posterior and basomedial amygdalar nuclei, LPO, nucleus pontis oralis (RPO), perifornical hypothalamus, periventricular, parafascicular, reuniens and mediodorsal (MDT) thalamic nuclei, centromedian, paracentral and centrolateral nuclei of the midline/intralaminar thalamus, suprachiasmatic nucleus (SCN), dopaminergic A13 region medial zona incerta (ZI), less VTA and SNc, retrorubral nucleus (RR), mesencephalic reticular formation (MRF), central linear raphe nucleus (Cli), anteromedial thalamic nucleus (AM) and LHb. MRN projects moderately to the restrosplenial (RSC, autobiographic memory), perirhinal, entorhinal, and frontal cortices, cingulate cortex, prelimbic, infralimbic, anterior insula (AI), but sparingly to remaining regions of the cortex (Vertes et al., 1999).
I propose (a new mechanism) of how ketamine and phencyclidine cause the loss of awareness and anesthesia by activating the 5-$HT_{2A}$ (or similar) receptors in the IPN. The activated IPN then activates MRN which by serotonergic projections activates claustrum. I claim that this IPN-MRN-claustrum activation induces cortical and some subcortical SWA and loss of awareness. Ketamine might also activate claustrum via 5-HT2a receptors. Kawai et al. (2022) found a large contribution of 5-HT2A receptors in IPN to the MRN-evoked IPN activation. I guess, that ketamine causes anesthesia by activating the 5-$HT_{2A}$ receptors on IPN, but analgesia by IPN→MRN→ventromedial PAG or caudal raphe nuclei pathways. My model would mean that MRN has an excitatory feedback effect on its feedforward input from the IPN. The feedback excitation of IPN by MRN might be fine, as the SWS phase takes around 60 minutes. So anesthetics and SWS-circuit induce loss of awareness by activating the MHb→IPN→ MRN→claustrum→cortex pathway that enables slow-wave oscillations, and loss of awareness. SWS promotes also sharp wave ripples, hippocampo-cortical replay, and cortical ripples activity.

## Loss of awareness induced by IPN→MRN→claustrum→cortical slow-waves.

I claim a new neural mechanism for inducing loss of awareness and anesthesia by activating the MHb→ IPN→ MRN→ serotonin→ claustrum→ cortical SWA (0.1 – 2Hz), especially in prefrontal cortex. The slow-wave activity is too slow to convey awareness and is usually found in slow-wave sleep. SWA in animals was also found during quiet wakefulness, especially during food consumption. Thus I propose that also in humans can be SWA found in wakefulness, during brief moments that cause slips of attention and awareness. The SWA might intrude awareness state especially after a meal or during boring repetitive tasks, which are not supposed to be recorded. The up-states of slow waves are coupled to hippocampal SWR and cortical/subcortical ripples that enable hippocampo-cortical memory replay. Theta state in wakefulness enables the encoding of new information but is also important in REM sleep, where dreams and restricted awareness occur. IPN and MRN inhibit theta state regions to avoid mixing the replayed with new information, and to achieve loss of awareness. The up-states of SWA enable memory replay, possibly to achieve strong potentiation of newly activated synapses.
I propose that loss of awareness is induced by several, or all anesthetics via partial or full activation of the MHb→IPN →MRN→ claustrum → prefrontal cortex SWA circuit. I proposed that mu-opioids act by activating the medial habenula, while ketamine and phencyclidine activate IPN and claustrum. Heightened activity in claustrum neurons that project to the anterior cingulate cortex (ACCp) corresponded to reduced sensory responsiveness during sleep (Atlan et al., 2024). They also found that their heightened activity correlated with disengagement and behavioral lapses, while their low activity correlated with hyper-engagement and impulsive errors.
I propose that the loss of awareness and brief lapses of attention in wakefulness are induced by activation of the human homolog of the rat infralimbic cortex, the vACC/a25 →IPN →MRN→ claustrum → prefrontal cortex SWA pathway. Cortical and especially prefrontal SWA, and inhibition of arousal, dopamine, NBM, and theta state circuits (by IPN-MRN) cause loss of awareness.
The Vila-Merkle et al. (2023) study results support my idea. They found that slow-wave activity of 0.5-1 Hz was inhibited by anxiogenic drug FG-7142 in male rats, but reappeared following infralimbic (vACC/a25 homolog) deep brain stimulation. Herkenham (1981) observed activated MHb and IPN after applying anesthetics pentobarbital, ether, and chloral hydrate in rats.

Dickey et al. (2022) found by intracranial recordings in humans that brief high-frequency oscillations (ripples) occur in the hippocampus and cortex and help organize memory recall and consolidation. They reported that these approximately 70 ms-duration, 90-Hz ripples often couple (within 500 ms), co-occur (less than 25-ms overlap), and have consistent phase lag between widely distributed focal cortical locations during both, sleep and waking, even between hemispheres. So cortical ripples occure both during recall and replay and re-activate the specific information/bits of relational, contextual, or event-based memories, that were encoded and linked/bound together in sequences of CA1 and CA3 neurons firing.
Other evidence in support of my claim is that Hunt et al. (2009) found that ketamine induces high-frequency oscillations around 180 Hz in NAc in quiet wakefulness in rats. I predict that ketamine activates through the 5-HT2a receptors the vACC/BA25 which then directly activates IPN. IPN projects to both dentate gyrus (DG→CA3→CA1→ SWRs →NAc ripples) and MRN (MRN→CA1 + claustrum→SWA in NAc), likely inducing hippocampal SWRs that activate fast ripples in cortex and NAc. The IPN via MRN serotonin activates the claustrum. The claustrum then induces cortical and subcortical SWA and a brief absence of awareness when the rat is in the safe consummatory stage.
I propose that an anesthetic dose of ketamine activates the 5-HT2A receptors in the IPN and claustrum. The IPN consequently activates the MRN that causes loss of awareness by activating the claustrum which induces cortical and some subcortical SWA. Claustrum has serotonergic afferents. So during anesthesia and SWS, serotonin from MRN activates the claustrum which induces cortical SWA. Narikiyo et al. (2020) found that in vivo optogenetic stimulation of claustrum in mouse induced a synchronized down-state (of SWA) featuring prolonged silencing of neural activity in all layers of many cortical areas, followed by a down-to-up state transition. In contrast, genetic ablation of claustral neurons attenuated slow-wave activity in the frontal cortex. Narikiyo et al. found that the mouse neocortex generates a large-scale SWA during sleep and awake rest. They induced the neocortical SWA optogenetically, by activating glutamatergic neurons in the claustrum. The generated Down-state caused a prolonged silencing of neural activity in all layers of many cortical areas, followed by a globally synchronized Down-to-Up state transition.
Yamaguchi et al. (2013) found that triangular septum (TS) projects to ventral MHb which projects to core IPN, while bed nucleus of the anterior commissure (BAC) projects to dorsal MHb which projects to peripheral IPN.

I propose that the SWA is induced by the DG→TS and BAC activation of the MHb→IPN→MRN→claustrum→cortical/subcortical SWA circuit in mammals, in SWS (including SWA in NAc). I claim that SWA can be induced in mice also in quiet wakefulness, by activation of the infralimbic cortex (vACC)→ IPN→MRN→claustrum→ SWA circuit. I suggest that vACC induces brief cortical SWA periods in quiet wakefulness also in humans, during less than second-long slips of attention (absentmindedness), possibly after a meal or during repetitive boring tasks. This wakeful SWA likely induces hippocampo-cortical memory replay and ripples.
Staines et al. (1988) found adenosine deaminase (ADA) immunoreactivity (IR) in the septofimbrial, triangular, and BAC nuclei neurons (of the postcommisural septum), and in their terminals in the medial habenula. These ADA-IR neurons projections (involved in adenosine catabolism and purinergic neuromodulation) joined the stria medularis, continued caudally within this fibre bundle, and terminated densely and focaly within the dorsal MHb. Fibers labelled for ADA terminated in substance P- containing subregions, while enkephalin-positive septohabenular fibres occupied distinct, non-overlapping dMHb subregions. Thus, I propose that the septofimbrial, triangular, and BAC nuclei induce SWS, SWA and loss of awareness also by activating dMHb via their adenosin or ATP release: TS + BAC + septofimbrial septum→ dMHb→ IPN→ MRN→ claustrum→ SWA.

I propose that the non-anesthetic, lower dose of well-being-improving ketamine acts by activating the vACC/a25 which then activates DRN serotonin. This serotonin inhibits GPi input to LHb, thus disinhibits the serotonin and dopamine release, increasing cognitive flexibility, working memory, and awareness.
Interesting is the role of RMTg in sleep, anesthesia and awareness. It is directly inhibited by mu-opioids in wakefulness - causing disinhibition of dopamine and DRN serotonin release. But RMTg is activated by mu-opioids in SWS and anesthesia via MHb-IPN-MRN-LPO-RMTg circuit - causing inhibition of movement, arousal, dopamine and DRN serotonin.

Alarm, worries, stress oppose safety, sleep, rest, growth, repair, and recovery states.

If the MHb and IPN are active, it is kind of safe to fall asleep, thus I claim that noradrenaline, dACC/prelimbic cortex, sympathetic system, dopamine, NI, PH, and LH inhibit the IPN system. Stress, novelty, and excitement hinder us from falling asleep. Familiarity, safety, opioids, cannabinoids, dentate gyrus input via the triangular septum to MHb, and anteromedial BNST/BAC input activate the MHb-IPN system. This SWS- and loss of awareness system inhibits the theta-promoting systems: MS/DBB, SUM, NI, MMB, nuclei of Gudden, SI, LDT, alertness, and REM sleep. Also septofimbrial nuclei project to MHb and IPN, so they might be activated by familiarity and known environments. If their afferents come from the hippocampal region (CA3 ?) that compares the incoming information and signals if the context is novel or familiar, then the familiar context might activate septofimbrial septum which computes familiarity as safe and activates the MHb-IPN system. In contrast, when we notice something new, interesting, or exciting, it is harder to fall asleep, than when we feel safe or bored. Molas et al. (2017) observed an increase in IPN dopamine signal during a social investigation of a novel but not familiar conspecific, and during exploration of the anxiogenic open arms of the elevated plus maze.

There is time to act (seek, work, think, learn, defend) and time to relax and sleep - linked to rest, safety, recovery, protheosynthesis, synaptogenesis, reconstruction, growth, and immune defense. I predict that inhibition of IPN and MRN by LHb (after pain, loss, worries, or disappointment) hinders and decreases the amount of restorative sleep and memory replay/consolidation, thus learning. Stress and fear (of negative actual or predicted events/harms) diminish memory replay and SWS sleep by inhibiting IPN and MHb: IPN by dACC/prelimbic (Beware!), noradrenaline (LC, danger, alarm), corticotropin-releasing hormone (CRH, stress) and LS (fear) input, and MHb by noradrenaline (LC and sympathetic system) and MS/vDBB (theta oscillations for new memory encoding). I predict that IPN mutually inhibits LS and medial LHb. I propose that the infralimbic/vACC input to IPN activates SWR and hippocampo-cortical/subcortical memory replay which strengthens the newly acquired memories/synapses in wakeful rest. For example during the consummatory state in rats, but also during quiet wake (no-novelty, no-recording) in humans in brief moments of attention lapses or unawareness (''switch offs'').

I found that an important calming effect on stress and enforcing effect on restorative SWS sleep comes from the vACC/infralimbic cortex (Vadovičová, 2024b). I propose that vACC activates IPN-MRN which inhibits the theta- and arousal-linked noradrenergic LC and relaxin-producing nucleus incertus (NI) regions, thus decreasing the alarm state of the brain and body. I predicted that the vACC also mutually inhibits the warning signal in the dACC/prelimbic cortex region (Vadovičová and Gasparotti, 2014). By activating the IPN, the vACC indirectly inhibits the LHb (overstimulated by bad outcomes and pain), thus disinhibiting serotonin release. I also showed how serotonin calms down stress, worries, and anger via dACC and amygdalar receptors.

Further evidence for the proposed role of the MHb-IPN circuit, is the loss of opioid efficacy in sleep-deprived individuals. It supports my ideas, that mu-opioids activate via MHb the analgesic and the anesthetic - loss of consciousness pathway that induces SWA and inhibits the theta states. In acute danger, the noradrenaline and sympathetic projections inhibit MHb, and the amBNST activation (active when it is 'safe to fall asleep') that activates MHb is attenuated by worries. Thus a higher mu-opioid dose is required to activate the MHb that is inhibited by worry and fear. Worries stop us from falling asleep and wake us up. Pain, loss, and negative outcomes activate LHb which inhibits MRN, DRN, VTA, and SNc, so these adversities diminish SWS also by suppressing MRN serotonin. Thus also depression, via overstimulated LHb inhibits MRN and SWS.

My MHb→IPN→MRN→ claustrum model proposed that amBNST/BAC, triangular, and septofimbrial septal nuclei, mu-opioids, and cannabinoids stimulate MHb, while the MS/vDBB, noradrenergic locus coeruleus (LC) and sympathetic system neurons attenuate the MHb output. Thus pain, worries, and thoughts about dangers and loss diminish the restorative processes of sleep, decrease SWS, the occurrence of ripples and slow-wave oscillations in the cortex, and relational memories replay and consolidation (Vadovičová, 2015).
On the contrary safety, satisfaction, and peace support the restorative functions of SWS by activating the MHb-IPN, and by increasing resilience to stress via vACC→DRN → serotonin-linked inhibition of LHb, amygdala, dACC, and noradrenaline. Lack of serotonin and overstimulated LHb diminish SWS in depression, and SWS attenuates LHb, depression, and anxiety via MHb→LHb suppression.
I suggest that nicotine consumption causes emotional relaxation and attenuates stress, anxiety, and worries by inhibiting lateral habenula. I propose that MHb inhibits LHb directly and polysynaptically via activating the MHb→IPN→MRN circuit. Evidence for this comes from a wide use of smoking as self-medication in anxiety states.

## Periacqueductal grey connectivity and OIRD.

The PAG receives cortical afferents from the insular, cingulate, prelimbic/dACC, infralimbic/vACC, motor, and premotor cortices (Torrealba and Müller, 1996). PAG might cause breathing depression by projections to PBN, NTS, or preBötC. The caudal NTS receives direct projections from the insular and infralimbic cortex (IL), PAG, CeA, LH, paraventricular hypothalamus (PVH), rostral NTS, and Kölliker-Fuse nucleus (Gasparini et al., 2020). The IL/vACC projections to NTS might transmit calming (parasympathetic) effects of safety and well-being on breathing and heart rate during wakefulness. The PAG, PBN, dorsal anterior cingulate cortex (dACC), posterior and anterior insula (AI) are parts of the nociceptive pathway. The lateral (lPAG) and dorsolateral (dlPAG) columns react in active coping strategies, such as fight or flight, while the ventrolateral (vlPAG) column is linked to passive coping, such as freezing (Bandler et al., 2000; Linnman et al., 2012; Faull et al., 2019).
Dopaminergic cell groups from area A13 medial zona incerta, A10 ventral PAG, posterior hypothalamic area, rostral linear raphe nucleus, A11 periventricular grey, and dorso-posterior hypothalamus project to dlPAG (Mesanvi et al., 2013).
So vPAG might be activated by hypercapnia and pain afferents from PBN, and release dopamine to dlPAG to induce alarm and arousal response to harm signals from PBN. I propose that RMTG in SWS and anesthesia inhibits vPAG, thus arousal or movement response. I predict that IPN inhibits dlPAG response to harmful interoceptive and sensory signals during IPN induced slow-wave sleep and mu-opiate anesthesia. Shah et al, (2005) found that morphine strongly decreased the pain-evoked activation in the dlPAG, amygdala, LH, and cingulate cortex (homolog of human PFC).

Trevizan-Baú et al., (2021) assessed the relative strength of the reciprocal connectivity of the predominantly lateral and ventrolateral PAG with core nuclei of the respiratory network using a retrograde tracer Cholera toxin B. They confirmed that the lPAG and vlPAG project to the Kölliker-Fuse nucleus (KFn), pre-Bötzinger complex (pre-BötC), and caudal raphé. They found the strongest projections to lPAG and vlPAG from the KFn and far weaker projections from the pre-BötC, Bötzinger complex (BötC), and caudal raphé. They suggested that reciprocal connectivity of the KFn and PAG has specific roles, most likely related to the upper airway patency regulation during vocalization or other volitional orofacial behaviors. So vlPAG seems to interrupt/suppress respiration during both freezing and vocalization, so it might also depress respiration by attenuating the pre-BötC or other respiratory nuclei. vlPAG to caudal raphé projections might attenuate respiration or pain processing in PBN. So mu-opioids evoked respiratory depression might lower respiration also via the MHb-IPN-vlPAG- KFn + pre-BötC pathway.

I guess that IPN can inhibit dlPAG, or activate vlPAG, or interact with other PAG subnuclei, leading to respiratory slow-down. So high dose of mu-opioids activates MHb which increases IPN output. IPN then attenuates dlPAG activation, causing lower arousal to noxious respiratory depression. IPN can via some (ventral?) PAG nuclei activate also analgesic pathway, as mu-opioids induce strong analgesia.
The PAG can modulate respiration by activating the nucleus of the solitary tract (NTS). The PAG has both, the afferents from, and efferents to NTS (Krohn et al., 2023). The pro-opiomelanocortin-expressing (POMC) neurons of the NTS project to the pre-Bötzinger complex and the cardiac vagal motor neurons in the nucleus ambiguus. Via these pathways, they can reduce inspiration and cardiac function, respectively (Krohn et al., 2023). Nucleus ambiguus controls also laryngeal muscles, and NTS controls via the phrenic nucleus the Diaphragm (Figure 5, Krohn et al., 2023). The NTS has bidirectional connections with most areas regulating respiration: the pattern generation areas, chemoreceptors areas, modulatory areas, limbic system, sensory regions, sensorimotor regions, premotor and motor areas (Figure 5, Krohn et al., 2023). Behbehani and Fields (1979) found evidence that an excitatory connection between the PAG and nucleus raphe magnus (NRM) and adjacent reticular formation mediates the stimulation-induced analgesia in NRM.
My mu-opioids →MHb→ IPN→vPAG→PBN + NTS + pre-Bötzinger complex + caudal raphe nuclei circuit claims seem to agree with the following Koo and Eikermann’s (2011) ideas: Normal respiratory rhythm generation is decreased by pre-Bötzinger complex inhibition. Chemosensitivity to hypercapnia and hypoxia are blunted by opioids at the retrotrapezoid nucleus, medullary raphe nucleus, and NTS. Opioids also decrease the central drive to the respiratory pump muscles and the upper airway dilator muscles. Opioid-induced respiratory depression can be reversed by naloxone. The respiratory drive is modulated by feedback from central and peripheral chemoreceptors and driven in a feedforward fashion by wake-active forebrain regions. I showed IPN and MRN inhibition of the arousal circuit. I proposed also OIRD via the IPN inhibition of DRN chemosensory output: MHb→IPN →down DRN →down lPBN.
Mu-opioids inhibit pain localy, at spinal, medular, and brainstem afferents, centraly via MHb-IPN circuit, and possibly also at sensory (PI) and affective (AI, dACC) cortical processing levels.

## Ketamine and psychedelics role in anesthesia, safety, and well-being.

As ketamine is used in anesthesia, the question is how it interacts with the MHb→IPN circuit and its PAG, MRN, DRN, and LPO effectors. And if it decreases or increases the respiratory depression? I propose two new mechanisms for anesthesia, loss of awareness, and antidepressant/anxiolytic effects of ketamine and psychedelics. I am the first to show how ketamine and 5-HT2a agents act via the 5-HT2a serotonergic receptors in the IPN, claustrum, and vACC/a25. A reduction in Rapid Eye Movement (REM) sleep was found after ketamine administration in humans on the day of the sleep study (Gottschlich et al., 2011). Serotonergic MRN projections inhibit theta oscillations (present in REM sleep and active wakefulness), as Kinney et al. (1995) and Vertes et al. (1994) produced hippocampal theta rhythm in the urethane anesthetized rat after injecting into the MRN either muscimol, procaine or serotonin 1A agonists. I propose that high ketamine doses, phencyclidin and nitrous oxide cause anesthetic effects by activating the 5-HT2a IPN and possibly claustrum receptors. The IPN then activates MRN which via serotonin release activates the claustrum and changes the hippocampal activity. Claustrum induces cortical and striatal slow-wave activity (SWA) linked to unawareness, SWS and anesthesia. 5-HT2a + Ketamine→ IPN→ MRN→ claustrum→ cortical SWA→ unawareness + anesthesia.

I propose that ketamine (low dose), psychedelics and nitrous oxide activate the 5-HT2a receptors in vACC/a25 (infralimbic cortex, IL in rats) and increase its well-being and safety signal. I made a new circuit-based model of WHY vACC activates serotonin release in DRN and HOW it interacts with the brain and mind in learning, choice behavior, feeling, and thinking. I stated that area vACC/a25, signals when humans gain safety, social and biological rewards, affection, nurture, and acceptance when we reach a comfort zone, our goals, win prospects, do something well, or make someone smile (Vadovičová and Gasparotti, 2013, 2014). I propose that lack of nurture or care in early life diminishes the brain-derived neural factor (BDNF) in vACC/a25, decreasing its neural growth. Lower vACC activation leads to lower DRN serotonin, followed by higher dACC, AI, and amygdala activity - more worries, fear, anger, and stress. So early life deprivations and fear change affective processing: danger→less well-being and safety→less vACC activation/signal→less DRN serotonin→less BDNF→less growth in vACC →more anxious, less stress resilient personality (higher dACC, CeA, and LHb strength/signal). The BDNF is known to increase synaptogenesis and neural growth. Chronic pain, worries, anhedonia, or trauma, overstimulate LHb and decrease serotonin release from DRN and MRN, and dopamine signal, causing anhedonia and depression. Dopamine moves us to seek, move, and make an effort for rewards, new, interesting, good - well-being, or survival-promoting things.

Mu-opioids activate the DRN serotonergic release (Welsch et al., 2023). I propose that mu-opioids increase the well-being and safety signal in the vACC/IL disynaptically, via its serotonergic receptors: mu-opioids→DRN serotonin →vACC up→CeA down. Shabel et al. (2012) found that input from the globus pallidus interna (GPi) to the LHb is excitatory, aversive, and suppressed by serotonin. I think that mu-opioids, ketamine, and psychedelics have fast antidepressant effects also by indirectly inhibiting, via serotonin, the LHb, anterior insula, dACC, D2 pathway of VS, GPi, and CeA. psychedelics + ketamine→ up vACC/a25→DRN serotonin→down LHb + AI + dACC + CeA. So they indirectly decrease the information processing about pain, loss, and bad outcomes, and disinhibit serotonin and (less) dopamine release. DRN serotonin decreases addiction, reward, and drug-seeking by lowering VTA dopamine release in the VS (thus attenuating D1 pathway) while disinhibiting dopamine release in the DS (thus strengthening D1 pathway).

I proposed that activation of vACC/IL by ketamine, serotonin, hallucinogens, or oxytocin increases well-being and safety feelings, and activates oxytocin release in the hypothalamus. Oxytocin increases feelings of being connected, is known to induce bonding and attenuate pain (i.e. during birth). Serotonin, oxytocin, and dopamine signaling are affected in autism (Vadovičová and Gasparotti, 2013). As a dopamine-inducing novelty, social rewards, and socializing are less valuable, joyful, and attractive in the autistic population. Their vACC growth and reward circuit connectivity might be weaker, decreasing serotonin and dopamine-linked learning.
I propose that vACC activation, ketamine, and hallucinogens (5-HT2a agents) increase cognitive flexibility by activating vACC→DRN serotonin release in the substantia nigra pars compacta (SNc), known to disinhibiting dopamine release. I predicted that specific medial frontopolar region processes, estimates, and signals the informational value and meaningfulness of things and events to us (Vadovičová and Gasparotti, 2013, 2014). This region codes the informational value, relevance, rightness, and validity (of things, options, ideas). This region/module is most strongly activated by the right, correct, relevant, and meaningful choices, predictions, and ideas. It is located superior to the rostral gyrus rectus, halfway between the vACC/a25 and the ventral medial frontal pole region that, in my opinion, encodes goals (and is strongly interconnected with prefrontal cortex (PFC) regions). I proposed that the information value and meanings processing region induces dopamine release in SNc to the cognitively valuable, relevant, true, right, meaningful, or important choices and information. This dopamine signal then reinforces the strength of linked (useful, interesting, relevant) neural representations in working memory, in PFC regions (DLPFC spatial processing, VLPFC objects-properties-interrelations-meanings, dorsomedial frontal pole (FP) temporal organization/context and event-based processing), and in the head of the caudate. Disinhibition of dopamine release from SNc by the DRN serotonin might be the cause of a sustained and elevated gamma power after sub-anesthetic ketamine. So ketamine might via SNc disinhibition (ket-vACC-DRN-serotonin-SNc) activate the nucleus basalis of Meynert (NBM), known for inducing fast cortical gamma oscillations, involved in attentive processing and coupling of cortical regions. So dopamine is important for awareness and is also released in dreaming during REM sleep.

The head of the caudate nucleus is part of the dorsal striatum and the cognitive cortico-striato-thalamo-cortical loop proposed by Alexander et al. (1986), involved in cognitive probabilistic selection and evaluation between options and course of actions, depending on the subject's experiences with them. So serotonin, ketamine, and psychedelics increase cognitive flexibility by disinhibiting the SNc dopamine. Value-based dopamine release in the head of caudate in the dorsal striatum increases a propensity to select cognitively optimal options and inhibits the affectively pleasant but short-term rewards followed by negative consequences (cakes, drugs, loss of health).

Dopamine enforces representations of potentially useful, new, interesting, relevant, valuable, or meaningful information in PFC, and the strength of D1 loop of basal ganglia. When deprived/hungry or hurt, serotonin drops → disinhibiting wanting, as we need to get rewards quickly to survive, despite the pain and obstacles involved. The thinking is narrowed to what is missing, cognition is perseverative and prevailing negative feedback inhibits us, via the D2 loop of the ventral striatum (VS). Enough serotonin in the brain makes us free/flexible to think about new and potentially good things to think about, hope for, pursue, or do, that are possible to win, instead of those that are not. So, the serotonin-evoked disinhibition of SNc dopamine decreases the affective bias for immediate rewards (selection via VS, VTA) and increases the cognitive bias (via enforced DS) in decision-making, by activating the D1 (right, optimal options) and inhibiting D2 (wrong options) loop of the cognitive striatum. Serotonin decreases impulsivity and motivation to seek/choose the most pleasant affectively good choice (via D1 loop of VS), by lowering dopamine release in VS, and decreases inhibitions/aversions and learned helplessness by inhibiting the D2 loop of VS (Vadovičová and Gasparotti, 2013, 2014). In addition, serotonin decreases fear and anger in amygdala and increases socializing.

MHb→IPN→MRN inhibits new memory encoding in hippocampus
and induces hippocampo-cortical memory replay state.

Ohara et al. (2013) used a trans-synaptic retrograde tracer and found disynaptic input to dorsal and ventral dentate gyrus (DG) from MHb, IPN, LS, endopiriform nucleus, PFC, and claustrum, and direct input from CA3, entorhinal cortex, MS/DBB, SUM, and few IPN and MRN neurons.
There was no or sparse retrosplenial cortex input to DG. Prelimbic and infralimbic cortex (alert/alarm vs safety) had disynaptic input to the ventral dentate gyrus, dACC→MS/DBB→vDG, vACC→ IPN→vDG, and via entorhinal cortex. They found cingulate cortex (human cognitive PFC homolog) disynaptic input to dorsal and ventral DG. The cortical amygdala projected directly to the rostromedial entorhinal cortex. Medial MHb and ventral LS reached the ventral DG disynaptically via dorsal IPN projections, while the lateral MHb via ventral IPN influenced the dorsal DG. They found topographically distributed disynaptic input pathways to dorsal and ventral DG from the MRN/lateromedial axis, IPN/ventrodorsal, MHb/lateromedial, caudal LS/dorsoventral, and claustrum/dorsoventral axis.
The firing fields of place cells in the hippocampus increase their dimensions from dorsal to ventral axis (Kjelstrup et al., 2008), and the spacing of firing peaks of single grid cells in the medial entorhinal cortex increases along a dorsoventral axis (Giocomo et al., 2007; Hafting et al., 2005). Interestingly, the IPN projects to entorhinal cortex, dentate gyrus, horizontal DBB, MRN, and VLPO. Thus, IPN can induce hippocampo-cortical memory replay state, and via hDBB ihibit gamma synchronizations linked to wake and REM sleep. IPN might activate the recent cognitive maps in the medial entorhinal cortex, that activate sparse dentate gyrus representations. DG might reactivate the CA1 assemblies, memory sequences producing ripples. CA1 project directly into cognitive prefrontal cortex, and affective mOFC, vACC and dACC regions, maybe in humans also to temporal poles conceptual representations, and to anterior insula. Event-based information processing and temporal organization of thoughts and plans is linked to dorsomedial frontal poles. They hold working memory of past and imagined events and interact with retrosplenial cortex autobiographic memory. Goals are likely encoded/processed in the region ventral to it. They influence and highlight information/representations in other prefrontal regions, whenever goals guide our behavior or thoughts. The IPN projects directly to DG and MRN, so by inducing serotonin release it changes the hippocampal state into a replay mode. IPN and MRN projections inhibit theta and gamma coupling linked to alert wakefulness and REM sleep, and inhibit arousal linked regions.
Slow-wave sleep is distinguished by slow, high amplitude EEG, while rapid eye movements co-ocure with REM sleep, dreaming, postural muscles atonia, and fast, low amplitude cortical oscillations which resemble wake state EEG (Aserinsky & Kleitman, 1953; Dement & Kleitman, 1957; Jouvet & Michel,1959). In SWS is BF cholinergic firing low, and firing rates in the hippocampus and multiple cortical areas are reduced. Cholinergic agonists in cortex switch the synchronized cortical SWA to asynchronous patterns characterizing wakefulness, and vice-versa cholinergic inhibitors induce cortical and hippocampal slow-waves waking animals **(**Jones, 2005). Basal forebrain (BF, NBM, SI, hDBB) induces gamma oscillations and long-range cortical coupling involved in awareness, arousal (Jones, 1995, 2003, 2005), new memory encoding, learning, selective attention (Chiba et al., 1999), and working memory (Teles-Grilo Ruivo et al., 2017). Also MS/vDBB that induces theta oscillations is part of BF. BF cholinergic neurons have greater activity during waking and REM sleep compared with SWS (Jones, 1993, 2003 and 2005; Lee et al., 2005; Xu et al., 2015). Using microdialysis in cortex and dorsal hippocampus, Marrosu et al. (1995) found that ACh release increased by 100% during quiet waking (QW) and by 175% during active waking compared to SWS baseline. During REM sleep resembled cortical ACh the QW levels, while hippocampal ACh rose to about 4-fold the SWS level or twice that of QW. They stated that the increased cortical ACh release reflects the desynchronized EEG of wakefulness and REM sleep, while the marked increase of hippocampal ACh during REM is related to the sustained theta activity there. In my opinion BF, dopamine and prefrontal cortex are important for consciousness, and in selection of bottom-up (i.e. by pain) and top-down information (by goals and strategic planning in frontal poles) that is interesting, important or relevant, has meanings. Being conscious helps us in moving around, making goals, questioning world, testing our guesses, learning from outcomes, planning. Hegedüs et al. (2023) made calcium recording of hDBB gabaergic parvalbumin projections on PV-expressing interneurons in medial septum, CA1 of the hippocampus, and retrosplenial cortex. Their study revealed prominent activation of these targets by air puff punishments. Further targets of hDBB parvalbumin neurons (BFPVNs) included the MS/vDBB, mammillary and supramammillary nuclei, paratenial thalamic nucleus, and frontal cortex. I suggest that IPN via hDBB interacts with these targets when inducing the hippocampo-cortical memory replay state and ripples linked to unawareness. The same group demonstrated that somatostatin projection interneurons have predominantly non-serotonergic input from MRN, and in wakefulness react to reward, punishment, surprising, and predictive cues. Momiyama and Záborszky's studies (2006) found that SOM interneurons presynaptically inhibit both GABA and glutamate release onto BF cholinergic neurons. SOM neurons inhibit Vglut2, cholinergic, and PV neurons and receive excitatory input from cholinergic and Vglut2 cells (Xu et al., 2015). Cholinergic neurons became active at the onset of running and licking and also in response to overt punishment regardless of behavioral context (Harrison et al., 2016). So movement excites both theta and gamma oscillations as it is part of new event (information bound in time and space: what with where and when) that are recorded by hippocampus.
The MRN is known to project to NBM, SI, SUM, MS/DBB, so it inhibits arousal, awareness, working memory, theta and gamma coupling, similar to the IPN efferents. Adrenaline, noradrenaline, dopamine, GABA, glycine, Vglut1, Vglut2, orexin, somatostatin (SOM), neuropeptide Y (NPY), substance P, and enkephalin synapses were identified on BF cholinergic neurons (Záborszky and Gombkoto, 2018). Neuropeptide Y and substance P are found also in MHb. Firing rate of cholinergic and parvalbumin-containing BF neurons increase during cortical wake and REM sleep, while neuropeptide Y neuron firing is accompanied by cortical slow waves (Duque et al., 2000). Local NPY neurons synapse on cholinergic neurons both in the SI/hDBB area (Záborszky et al., 2009) and in the caudal globus pallidus/SI area (Nelson and Mooney, 2016). NPY, via NPY Y1 receptors, inhibits the majority of cholinergic neurons. (Záborszky et al., 2009).

I propose that anesthetic doses of ketamine, phencyclidine, or nitrous oxide (N2O) activate via IPN 5-HT2a receptors the MRN serotonin release. MRN serotonin then inhibits both theta states: alert wakefulness and REM sleep, by inhibiting the theta oscillations promoting circuit (SUM, MS/DBB, NI, LDT, nuclei of Gudden..). I propose that MRN serotonin activates the cortical slow-wave oscillations (SWA) by activating the claustrum. Serotonin interacts with basal forebrain, where it inhibits the awareness linked gamma oscillations and evokes replay and ripples linked coupling that that is brief. The circuit involved in SWS, cortical/subcortical SWA, hippocampo-cortical memory replay (of information bound together by space, time, interrelations and meanings), and unawareness includes: DG + septofimbrial nucleus → posterior septum. Posterior septum + amBNST/BAC → MHb → IPN → MRN → serotonin → hippocampus + claustrum → SWA.
The triangular septum might activate the MHb when the dentate gyrus recording capacity for the new event-based and relational information is full, and interference would degrade the encoding (Vadovičová, 2015). Septofimbrial septum (input from CA3?) might activate MHb and sleep by familiarity input about the environment (linked to safety). Boring, non-novel data might induce sleep too, but nice exciting new environment my hinder humans and rats to fall asleep even if tired (safety reasons). I guess that the MRN serotonin interacts with the hippocampus and other regions to suppress the theta and gamma oscillations linked memory encoding, to not interfere with the hippocampal memory replay in CA1, sharp wave ripples, and memory consolidation.
MRN serotonin and gabaergic projections interact also with the thalamic nuclei and retrosplenial cortex processing (autobiographic memory). Their interactions with hallucinogens (via 5-HT 2A receptors) might cause sensory hallucinations. The MRN serotonin projections to the hippocampus are in my opinion important for the enhancement of protein synthesis, synaptogenesis, and dentate gyrus neurogenesis during SWS. I think that serotonin induces BDNF production, in rest and SWS.

## MHb-IPN effectors in Respiratory depression (OIRD).

Hypnotics and opioids decrease respiratory drive. Eikermann et al. (2012) found that ketamine activates breathing and abolishes the coupling between loss of consciousness and upper airway dilator muscle dysfunction in a wide dose range. He found that ketamine compared with propofol, has 1,5-to-2-fold higher values of inspiratory time and might help stabilize airway patency during sedation and anesthesia. I suggest that a non-anesthetic dose of ketamine increases respiration by disynaptically activating DRN, counteracting the inhibitory effect of mu-opioids and other MHb-IPN activating anesthetics on DRN chemosensation: ketamin→vACC →up DRN and hypercapnia induced increase in respiration. Low doses of mu-opioids
directly activate DRN, but higher doses inhibit it via MHb-IPN:
MOR →up MHb→up IPN→down DRN and its chemoreceptive output to ventilation regions.
up MHb→IPN→ MRN→up LPO→up RMTg→down DRN + dopaminergic vPAG/A10 + other dopaminergic nuclei.
DRN is one of the wakefulness-inducing regions, so its indirect activation by ketamine (via vACC/a25) might increase the chance of waking up during hypercapnia. Still, the question is if ketamine, via serotonin receptors in IPN, interacts with the part of the MHb→IPN→PAG circuit linked to respiratory depression. It depends on the specific IPN projections from the ketamine activated IPN subregions. If they polysynaptically strongly depress respiratory nuclei or not.

## GLP-1RAs interaction with MHb-IPN system.

I propose a neural mechanism for glucagon-like peptide-1-receptor agonists (GLP-1RAs) induced neurogenesis, anti-inflammatory effects, and protective effects against atheroma plaques by activating the medial habenula and restorative functions of the SWS. The MHb then interacts with the immune system and microglia, and through its IPN→MRN effectors activates the MRN serotonin. Serotonin (usually released during restorative SWS) activates BDNF brain-derived neural factor, dentate gyrus neurogenesis, decreases prostaglandin-linked inflammations, and protects against endothelial factor-linked inflammations in the cardiovascular system. On the contrary, chronic noradrenaline and CRH inhibit the MHb-IPN-MRN linked recovery and regeneration state, thus the body is more vulnerable to neurodegeneration and oxidative harm.

Clinical and experimental research shows that GLP-1RAs reduce inflammation and pro-inflammatory cytokines in neurological disorders, inflammatory bowel disease, and diabetic complications, modulate immune cell signaling, decrease glucose and prostaglandin in blood, and have the anti-inflammatory benefits beyond their recognized function in blood sugar regulation and weight control (Alharbi, 2024). GLP-1 and its receptor agonists influence thermogenesis, blood pressure, neurogenesis, neurodegeneration, retinal repair, and energy balance (Rowlands *et al.*, 2018). The GLP-1 and its transcriptional effects are cardioprotective, by decreasing the formation of atheroma plaques (Patel *et al.*, 2018). Semaglutide, used in type 2 diabetes treatments, is a glucagon-like peptide-1-receptor agonist, which causes delayed gastric emptying, reduced appetite, and weight loss. Semaglutide has cardioprotective properties in the mouse (Hinnen, 2017). Chronic inflammation can cause insulin resistance, so MHb activation and SWS-linked anti-inflammatory effects might delay diabetes type 2.

I propose that GLP-1RAs decrease addictions, withdrawal, and stress, and increase safety and well-being by activating the MHb-IPN-MRN-serotonin circuit. MHb and IPN inhibit VTA dopamine, and LC noradrenaline, though in SWS, it might still decrease craving and stress. Previously I claimed that MHb inhibits LHb, giving a relief (feeling lighter) to depression and anxiety. Also IPN, and MRN serotonin inhibit LHb. So by activating medial habenula, the GLP-1RAs decrease lateral habenula overstimulation, caused by chronic stress, anxiety, and withdrawal. Activation of MHb gives some brake from LHb overstimulation, increases SWS, and decreases noradrenaline and stress.
As GLP-1RAs increase feeling of satiety, I suggest that they also activate vACC (infralimbic cortex in rodents), the well-being and good range of homeostasis (comfort zone) signaling region. When you have enough, you don't need more. Activation of the vACC would directly activate satiety, fulfillment, feeling fine, at ease, in peace, ready to socialize and think freely (without feeling deprived of basic biological needs such as food, affection). vACC activates DRN serotonin release, thus vACC activation by GLP-1RAs would attenuate the affective choice of immediate rewards through decreasing the output of D1 direct pathway of ventral striatum, but increase the output of D1 pathway of dorsal striatum linked to cognitive and motor selection, including reasoning and goal-directed longterm choice behavior. Further, serotonin activation (via vACC to DRN projections) would decrease learned helplessness, affective inhibitory avoidance and aversion by attenuating the D2 indirect pathway in ventral striatum. In addition I propose that vACC via PAG and PBN attenuates the pain pathway. The vACC activation also increases cognitive flexibility by increasing disinhibiting dopamine release in prefrontal cortex, and by the fact that non-deprived persons have time/freedom to solve other problems and joys, instead of chasing the one being deprived of.

## Inhibition between anesthetic circuit and arousal evoking dopamine + noradrenalin.

Hypercapnia leads to increased activity in the locus coeruleus (LC) which is an evolutionary conserved reaction observed in amphibians and mammals (Elam *et al.*, 1981). I suggest that inhibition of MHb-IPN by noradrenaline and dopamine agents can interrupt respiratory and heart rate depression evoked by mu-opiates, causing arousal and awakening from anesthesia and SWS. Molas et al. (2023) used viral synaptic tracing in the dopamine transporter-Cre mouse line. They found that dopaminergic VTA neurons project to the nucleus accumbens shell and the caudal IPN. They observed an increase in IPN dopamine signal during the social investigation of a novel but not familiar conspecific and exploration of the anxiogenic open arms of the elevated plus maze. I claim that novelty, rewards, and interesting information induce dopamine and arousal, which decreases our chance of falling asleep, by inhibiting the MHb-IPN circuit. So the novel environment, people, and informational novelty activate dopamine which inhibits IPN and activates the D1 direct striatal pathway to increase motivation to seek and pursue valuable choices and environments longer. To avoid harm, the amygdala responds to threats and activates dopamine that moves us into active avoidance, self-defense, anger, or violence, and dopamine reciprocally activates/fuels the amygdala (Vadovičová and Gasparotti, 2014, 2013). Solt and team (2014) proved that electrical stimulation of the VTA induces reanimation from general anesthesia.

## MHb role in Sudden Infant Death Syndrome.

Another line of evidence supporting the MHb role in inducing respiratory depression and SWS, is linked to nicotine. High prenatal exposure to nicotine in humans and mammals leads to depressed respiration in early postnatal life. Nicotinic acetylcholine receptors were found in MHb and IPN (Salas et al., 2004). High prenatal exposure to nicotine might attenuate arousal-evoking regions by activating the MHb-IPN-MRN circuit. It might also enhance the MHb-IPN output to DRN and specific PAG nuclei. My model claims that IPN attenuates the ventilation and arousal responses to hypercapnia by inhibiting the hypercapnia sensing DRN neurons (in SWS and anesthesia) which via serotonin activate 5-HT2a receptors in the lateral PBN, and 5-HT4 receptors in the preBötzinger Complex. DRN has chemosensors for hypercapnia. Kaur et al. (2020) proved that DRN releases serotonin that via 5-HT2a receptors activates dorsal lateral PBN neurons that induce arousal. Lie et al. found that activation of neurons expressing the μ-opioid receptors in the lateral parabrachial nucleus of the pontine respiratory group increases respiration, and shows tight correlation with respiratory rate in mice. Inhibition of these neurons by morphine injection induced respiratory depression, and their activation by several excitatory G protein–coupled receptor agonists (TCB-2 a high affinity 5-$HT_{2A}$ receptor agonist, CCK8S, and Substance P) increased the respiratory rate and rescued mice from OIRD. These same drugs have been shown to modulate respiratory rhythm by activating other respiratory centers (Niebert et al., 2011; Ellenberger and Smith,1999; Hedner et al., 2011). Also the 5-HT4a agents were found to activate respiration and rescue OIRD. Manzke et al. reported that serotonin 5-$HT_{4(a)}$ receptors are strongly expressed in respiratory pre-BötC neurons and that their selective activation protects spontaneous respiratory activity. Treatment of rats with a 5-$HT_4$ agonist overcame fentanyl-induced respiratory depression and reestablished stable respiratory rhythm without loss of fentanyl's analgesic effect.

## Immune and restorative functions interact with the MHb-IPN-MRN circuit.

Disruptions of the MHb-IPN circuit decrease the MRN's serotonin release, the amount of SWS, and the restorative functions of SWS. Lack of serotonin and SWS causes more stress, harmful neuroinflammation, and neurodegenerations (e.g. more oxidative stress and autoimmune reactions) but less BDNF production, recovery, regeneration, synaptogenesis, and protective immunoreactivity. The MHb interacts with the immune system and microglia and possibly potentiates the immune response to bacterial and viral lipopolysaccharides, the anti-inflammatory and anti-tumor effects. I think that MHb induces restorative fever-linked sleep which can last days in influenza illness. Post-mortem study of brain sections of patients with depression found reduced volume, cell number, and mean cell areas in MHb (Ranft et al., 2010).
Sugama and his team found high interleukine-18 expression in MHb. The MHb also activates melatonin production in the pineal gland, which might increase regenerations and decrease neurodegenerations. Melatonin is inhibited by light and NA. Stress and noradrenaline inhibit both the MHb-IPN-MRN circuit and SWS, and lead to higher levels of prostaglandins and harmful inflammation. This also works the other way round, serotonin and its activators ketamine and hallucinogens attenuate noradrenaline and stress and activate the calming effects of the serotonin pathway. Inhibition or lesion of dorsal MHb in zebrafish elevated freezing behavior, anxiety, and alarm substance secretion (Agetsuma et al., 2010; Lee et al., 2010; Mathuru and Jesuthasan, 2013). IL-18 is expressed in rat brain in the cerebellum, hippocampus, hypothalamus, cortex, and striatum in astrocytes and microglia (Culhane et al., 1998). I suggest that IL-18 from the MHb activates via the IPN→MRN→claustrum + MS/DBB + SUM pathway the microbial infection-induced sleep response. Kubota et al. (2002) found that IL-18 is involved in the sleep responses to infection. They showed that higher doses of IL-18 markedly increased NREM/SWS sleep and brain temperature (also during the active period), and anti-IL-18 antibody attenuated the bacterial peptidoglycans-induced sleep in rats and rabbits. Their IL-18 injection during the light period also increased NREMS. Interestingly peptidoglucans from mushrooms have anticancerogenic effects by activating immune response against tumors and by decreasing metastases through decreasing angiogenesis, so they likely activate the IL-18. IL-18, a monocyte-derived cytokine enhances local antitumor immune responses by: activating natural killer cells and T cells (Kohno et al., 1997), inhibiting angiogenesis synergistically with IL-12 (Coughlin et al., 1998), inducing apoptosis in tumor cells (Hashimoto et al., 1999), mouse cancer cells growth inhibition (Tamura et al., 2003). Prevention of colon cancer spread is linked with elevated serum IL-18 levels (Goto et al., 2002). Takashi et al. (2006) found that cimetidine behaved as a partial agonist for histamine H2R (showing 35% agonist activity compared to histamine) and activated caspase-1 in monocytes. Both histamine and cimetidine concentration-dependently induced the production of IL-18 in monocytes via H2R, elevating cAMP that activates protein kinase A (PKA). Cimetidine treatment inhibits histamine-initiated angiogenesis via reducing vascular endothelial growth factor expression (Gifford and Tirberg, 1987). Perhaps this histamine-induced endothelial growth helps to seal the wounds that occurred in wakefulness but increases angiogenesis-linked tumor growth and embolias. Targeting the intratumoral dendritic cells can induce antitumor immunity (Furumoto et al., 2004). Histamine through H2 receptors inhibits the production of interleukin-12 (van der Pouw Kraan et al., 1998) but also protects mice from P. acnes-primed and LPS-induced hepatitis (Yokoyama et al., 2004). Also, adrenaline and NA agonists inhibit via β2-adrenergic receptors the IL-12 production but induce the IL-18 production which induces TNF-α and interferon IFN-γ, and inhibits IL-10 production in human peripheral blood monocytes (Takahashi et al., 2004). So, the stress-induced sympathetic system activates IL-18 in peripheral blood monocytes, peptidoglycans activate IL-18 and the MHb-IPN-MRN linked SWS/NREM sleep, and stress inhibits the MHb-IPN output and NREM sleep. How will NA and adrenaline influence the IL-18 synthesis by MHb?

Central or systemic administration of IL-1β or tumor necrosis factor (TNF)-α increases the NREM sleep in several species (Krueger et al., 1984) and activates the transcription factor nuclear factor kappa B (NF-κB), (Ballou et al., 1996) which promotes production of sleep regulating nitric oxide synthase (NOS), cyclooxygenase-2 (COX-2), nerve growth factor, adenosine A1 receptor, IL-2, IL-1, and TNF. IL-18 stimulates TNF-α production which induces IL-1β (Puren et al., 1998). These 3 substances activate NF-κB (Matsumoto et al., 1997; Robinson et al., 1997). Adenosine is produced in the ventral striatum during wakefulness and indirectly inhibits the VTA dopamine at the end of the day, thus also arousal and effort, making us sleepy and inhibited.

Sugama and Kakinuma (2020) concluded that NA modulates microglial activation in MHb in stress response, thereby determining their responses ranging from resting to activation state depending on the host's stress level or whether the host is awake or asleep. They stated that microglia under resting conditions may have constructive roles in surveillance, such as debris clearance, synaptic monitoring, pruning, and remodeling. In contrast, once activated, microglia become less efficient in brain waste clearance, and instead via cytokine or superoxide release increase neuroinflammation. IL-18 is a proinflammatory cytokine expressed in a variety of tissues including immune cells and superior MHb, and its production could be induced in the adrenal gland following activation of the hypothalamus-pituitary-adrenal axis (Sugama and Conti, 2008). The MHbS neurons interact with the immune system and stress because they increase IL-18 production in response to restraint stress (Sugama et al. 2002). The IL-18-expressing neurons and noradrenergic fibers are co-localized in the MHbS, so Aizawa et al. (2012) suggested that IL-18 production might be under the control of the noradrenergic inputs. They reasoned that the IL-18 in MHbS might modulate efficiency in the habenulo-interpeduncular transmission according to the central noradrenergic activation under stressful conditions. The adjacent MHbCd subnucleus expresses both IL-18 and Substance P. Interestingly they found mu-opioid receptors expression only in the lateral part of the MHb (MHbVL). Substance P is also released in a few pain-modulating brainstem pathways (in PBN and vPAG). So, I wonder if the substance P-producing MHb subnucleus forms a functional analgesic pathway with brainstem and spinal nuclei with the substance P receptors. Interestingly, when injected intracerebrally into mice, human leukocyte, but not fibroblast or immune interferon, caused potent endorphin-like opioid effects (Blalock and Smith, 1981). These effects included analgesia, lack of spontaneous locomotion, and catalepsy. All of these actions of human leukocyte interferon were preventable and reversible by the opiate antagonist naloxone.

## Discussion: Limitations and Scope of the Proposed Model

While the MHb-IPN model offers a integrative explanation for a range of neural phenomena, several limitations and necessary boundaries of its scope must be acknowledged.

First, the model's strength and also a potential weakness is in attributing the regulation of awareness, memory replay, and respiration to a single MHb-IPN circuit that is proposed to induce slow-wave sleep. Its primary function might be induction of a slow-wave dominant state, with other effects being secondary consequences of this global brain-state shift. This model does not seek to replace established systems, such as the thalamocortical loop for spindle/SWA generation or the VLPO for sleep maintenance, but rather proposes an upstream initiator and integrator that interacts with them.

Second, the causal claims made herein are, at this stage, theoretical predictions. While based on correlative evidence (e.g., metabolic imaging), direct interventional evidence is advised to confirm causality. The language of 'activation' and 'inhibition' should be interpreted as postulated mechanisms within the model's framework.

Third, the model currently simplifies the neurotransmission within these circuits. The MHb, IPN, and MRN are heterogenous nuclei with complex neurochemistry. Future refinements will be necessary to incorporate the specific roles of GABAergic, glutamatergic, cholinergic, substance P, or neuropeptid Y subpopulations within these regions. Or to disentangle the IPN-analgesic from the IPN-OIRD pathways.

Finally, the model's applicability to human physiology can be proved through neuroimaging and clinical observation.

## Discussion: Comparative Analysis with Competing Theories.

My model integrates and complements rather than replaces established frameworks, such as the VLPO Sleep Circuit, or GABAergic effects of anesthesia:

- VLPO maintains sleep state via sustained GABAergic inhibition of arousal nuclei
- IPN-MRN neurons initiate sleep transition by inducing cortical SWA, and GABAergic (+ serotonergic) inhibition of arousal, theta and gamma oscillation promoting nuclei. IPN activation includes inhibition of the G-protein receptors coupled K channels. Many Anesthetics activate the IPN receptors or their signaling cascade.
- Proposed interaction: MHb-IPN system activates VLPO directly or via MRN activation.

Relationship with Thalamocortical Theories:

- Thalamus generates and propagates SWA, but requires cortical drive

• Claustrum, activated by IPN→MRN circuit, provides this initial cortical synchronization
• Thus, this pathway might provide the 'ignition switch' for thalamocortical oscillations

• IPN might synchronize generation of spindles by TRN, through activating the acetylcholinergic and GABAergic hDBB projections: IPN→hDBB→TRN (thalamic reticular nucleus)
This integrative perspective positions the MHb-IPN circuit as a state-transition initiator working with established maintenance systems.

Acetylcholine in cortex, hippocampus, and arousal nuclei, released from NBM, SI, hDBB, MS/vDBB, and LDT/PPT, promotes memory encoding, theta-gamma coupling, and awareness. Surprisingly, acetylcholine and nicotine in MHb-IPN system induce SWS with memory replay – thus a mutually antagonistic brain state. My model predicts that cholinergic IPN activates GABAergic IPN projections that inhibit SUM, MS/vDBB, NI, VTg, DTg, LDT, LHb, LH, LC, LS, vPAG, DRN, and wake-on hDBB populations. MRN inhibits also SI, linked to awareness.

Proposed neural mechanisms show HOW activation of MHb by mu-opioids, nicotine, acetylcholine, ATP (and adenosine?), glutamate, substance P, GLP-1RA, interleukin 18, and disinhibition of MHb by cannabinoids activates IPN and SWS. IPN then activates MRN that via serotonin activates claustrum that induces SWA.
My model predicted the role of the dentate gyrus→triangular septum→ MHb→ IPN→ MRN→ hippocampus + claustrum→ cortical SWA circuit in the slow-wave sleep, anesthesia, loss of awareness, hippocampo-cortical replay and consolidation of event-based and hippocampally bound memories (linked by space, time, and interrelations), sharp wave ripples, cortical ripples, respiratory depression, anxiolytic effect, rest, recovery, repair, serotonin- and BDNF-linked neural growth and synaptogenesis. There is mutually reciprocal inhibition between the MHb-IPN-MRN system activation and the theta states: alert wakefulness and REM sleep, new memories encoding, arousal, and activation of theta states promoting regions: MS/DBB, SUM, NI, NBM, SI, hDBB, LDT, PPT, LH orexin, PH histamine, LC noradrenalin, and VTA dopamine. MHb-IPN-MRN circuit inhibits through its LPO→RMTG projections during SWS also the dopaminergic VTA, SNc, ZI/A13, vPAG, and possibly other dopaminergic nuclei.
MRN and IPN projections to the hippocampus and claustrum induce hippocampal memory replay mode and cortical SWA, respectively. Both promote cortical ripples at the up-state of SWA.
The MRN's serotonin activates BDNF, neural growth, and synaptogenesis in the hippocampus during SWS. I proposed that the mu-opioids, nicotine, endocannabinoids, some anesthetics, and safety activate MHb which activate IPN, while ketamine and phencyclidine via 5-HT2a activate IPN. On the contrary, dopamine, noradrenaline from locus coeruleus and sympathetic system input, and GABAergic input from MS/vDBB inhibit MHb-IPN system.
Zhang et al. (2024) found that the systemic administration of dual inhibitor of cannabinoid degradation promotes contextual fear memory consolidation and recall, enhances hippocampal activity, and its anxiolytic effects depend on CB1 receptors. My slow-wave sleep model showes why. That endocannabinoids by disinhibiting MHb strengthen hippocampal contextual memory consolidation, by MHb→LHb inhibition cause anxiolytic and antidepressant effects, and by their MHb/SWS/microglia interactions attenuate neuroinflammation and neurodegeneration.
Hashem et al. (2021) found that inhibition of cannabinoid degradation (strengthens 2-AG→CB1/2 receptor signaling) attenuates reactivity of microglial cells and astrocytes (in the cortex and hippocampus), reduces proinflammatory cytokines, astrogliosis, phosphorylated GSK3β and tau, and alleviates tauopathies, accumulation and deposition of intracellular hyperphosphorylated tau protein in transgenic tau mouse model of Alzheimer disease.

This study described new circuits for opioids-evoked respiratory depression:

mu-opioids →MHb→IPN → inhibits DRN → less serotonin → less lateral PBN + pre-BötC.
mu-opioids →MHb→IPN → PAG → PBN + pre-BötC + NTS.
The MHb and IPN regions were never before linked to respiratory depression/slowdown or loss of awareness. Science literature up to date refers only to the MHb-IPN role in aversion, stress, and anxiety. This study explained how danger, stress, and anxiety mutually inhibit slow-wave sleep, loss of awareness, analgesia, anesthesia, recovery, immune defense, and regeneration, by inhibiting the MHb-IPN circuit. So my findings extend and complement the current understanding of the medial habenula role.

My novel finding is that the IPN can inhibit DRN directly and also indirectly via the IPN→MRN→ LPO→RMTg→ DRN projections. I think respiratory slowdown is a normal function of slow-wave sleep, which decreases oxygen consumption, when the body is not moving. But, unnaturally high doses of mu-opioids can cause respiratory depression and death.
The overstimulation of the MHb-IPN-DRN down- circuit by strong fetal exposure to nicotin, likely causes ventilatory deficits in mammalian neonates, known as sudden infant death syndrome (SIDS), linked to high fetal nicotine intake.

Optogenetic activation of the MHb can test my idea that the MOR and some anesthetics activate the MHb→IPN→ MRN→ serotonin→claustrum →cortical/subcortical SWA circuit. These agents can through slow-wave activity (SWA) in cortical and striatal neurons induce loss of awareness. I predicted that optogenetic mu-opioid receptors (MOR) activation, specifically in MHb will cause opioid-induced respiratory depression (OIRD). High morphine dose in MHb would cause respiratory arrest. I predicted that several anesthetics activate MHb or IPN. I suggest that 5-HT2a, ketamine, phencyclidine and nitrous oxide administration into IPN, or its optogenetic activation will activate the MRN→ claustrum pathway and cause cortical SWA, loss of awareness and working memory. Important factor for loss of consciousness in SWS and anesthesia is the suppression of dopamine burst firing by the MHb→IPN effectors, and inhibition of working memory. Both dopamine and noradrenaline can evoke theta (via MS/DBB) and gamma (via basal forebrain) synchronizations of cortical/subcortical regions. towards attention provoking, danger predicting, reward predicting, interesting, novel, or meaningful information.

I proposed that adenosine activates slow-wave sleep, SWA and loss of awareness beside other neuronal targets, also by activating the MHb-IPN-MRN-claustrum-cortical SWA pathway at MHb receptors. I suggest that substance P from the MHb might via the IPN-PAG circuit attenuate afferent pain pathway. I suggest that substance P might activate a subset of cortical and striatal nNOS1/SSH interneurons known to fire during SWA. They might promote high frequency ripple oscillations during hippocampo-cortical memory replay.
This work found how MHb-IPN-MRN system induces cortical and subcortical SWA, loss of awareness, memory replay, and anesthesia. I proposed that IPN induces SWR by projections to DG, entorhinal cortex, hDBB, and MRN→ claustrum→ cortex. I claimed that IPN inhibits those hDBB projection interneurons (main parvalbumine and somatostatin populations) that are active in aware processing, learning, working memory and memory encoding. But that IPN activates smaller subset of hDBB gabaergic projection neurons that enable subcortical and cortical ripples - brief (120-200 Hz) oscillations during hippocampo-cortical replay (in UP states of SWA ). I proposed that MRN activates SWA, SWS, and anesthesia, claustrum and subset of SWS active basal forebrain neurons (NBM, SI) e.g. the neuropeptide Y neurons. The MRN inhibits MS/vDBB via 5-HT2C receptors, causing theta rhythm suppression (Hajos et al., 2003). I claimed that MRN and IPN inhibit coupling of theta and gamma oscillations (via basal forebrain and claustrum) to SEPARATE the memory encoding and replay states, to not mix the unrelated facts/information together. They promotes memory replay state, SWA, ripples, and loss of awareness.

I proposed new neural mechanisms for ketamine effects on the mind and brain. Ketamine induces anesthesia, safety, and well-being by activating the 5-HT2a receptors in the IPN, claustrum, and vACC/a25. A high, anesthetic dose of ketamine activates the IPN via its 5-HT2a receptors. The IPN then activates MRN serotonin and GABA release. The serotonin and ketamine activate via 5-HT2a receptors claustrum, which induces SWA in cortical and some subcortical regions (NAc). The SWA is too slow to convey awareness, but at its up-states occur hippocampo-cortical memory replay, with SWRs (fast, 150-200 Hz oscillations in hippocampus and some subcortical regions, NAc) and cortical ripples. MRN inhibits the recording of new information, by inhibiting the theta states, alert wakefulness, arousal, and REM sleep-promoting regions. Ketamine→ IPN→ MRN→ claustrum→ SWA in cortex→ anesthesia + unawareness.

Lower ketamine and hallucinogen doses activate serotonergic 5-HT2a receptors in vACC/IL. The activated vACC/a25 induces serotonin release in DRN which again activates vACC. Serotonin induces BDNF and synaptogenesis in vACC. Serotonin and vACC directly calm fear and stress by inhibiting the anterior insula (affective pain and negative feelings, processing bad qualities of things, persons, and conduct), central amygdala CeA (threat processing, anger, self-defense, violence), and dACC/prelimbic cortex (warnings signal and worries about current or predicted loss and pain). Serotonin inhibits both LHb and its main input from dACC and AI. Thus vACC/BA25 activation and the induced serotonin release have fast antidepressant and anxiolytic effects in the brain.
Ketamine→vACC→DRN serotonin→down AI + dACC + LHb + CeA→anxiolytic +antidepressant
Ketamine →vACC→down AI + dACC + CeA →less LHb activation→anxiolytic + antidepressant
Ketamine →vACC→DRN serotonin→SNc disinhibition→dopamine in PFC→cognitive flexibility
I proposed that vACC/IL activation attenuates pain via its PAG and PBN projections, and also causes the placebo effect. Believing that something helps or seeing a glass half-full activates vACC which induces DRN serotonin release, strengthening resilience to pain, loss, and despair. I proposed activation of analgesic IPN→PAG and IPN→MRN→ magnocellular raphe nucleus pathways through both the MHb→IPN and vACC/a25→IPN projections.
Claustrum might differentialy recruit the specific cortical interneuron subtypes and suppress those that are active in attention, memory encoding, and theta-gamma coupling, such as parvalbumine and VIP synthetising interneurons. There are specific claustral and basal forebrain inputs to the somatostatine-, neuropeptide Y-positive, nNOS1/substance P receptors-interneurons subgroups in the cortex. If serotonergic MRN projects to medial prefrontal cortex nNOS1 interneurons in rodents, does it activates them, promoting cortical ripples (or SWA)? How the nNOS1/SSH discriminates serotonin signal from DRN versus MRN? As they have opposite effects on sleep (SWS) and arousal. Ripples in rat mPFC and in subregion of human medial frontal poles might potentiate synapses between neurons (assemblies) that represent and link event-based memories, and relationaly linked information (i.e. task/sets, plans, ideas), bound to temporal, spatial or relational context. The seminal work of McCormick & Bal (1997) on Thalamic Reticular Nucleus (TRN) and Spindles revealed fundamental mechanisms by which the TRN generates synchronized thalamocortical oscillations, including sleep spindles, via its reciprocal connections with thalamocortical neurons. It describes the core circuit architecture that gates/suppress sensory input to cortex and promotes memory replay. My model suggests that the MHb→ IPN→ MRN→ claustrum→ cortical + striatal slow-waves pathway provides the "cortical drive" that enables the synchronized cortico-striato-thalamo-cortical oscillations described in their findings. Also the IPN→ hDBB→ TRN pathway might synchronize the TRN, while the MRN might (besides the known MS/vDBB and theta) inhibit also SI or NBM (arousal, attention, cognition). IPN likely decreases motor output and awakening also by inhibiting the central medial, parafascicular, and central lateral thalamic nuclei. Halassa et al. (2014) demonstrated that the TRN contains functionally distinct subnetworks that are regulated according to sleep and attention states. They found that sensory thalamus-projecting TRN neurons participate in spindles, increase activation with spindle power, synchronize with slow waves during sleep, are suppressed by attentional states, and their optogenetic stimulation versus inhibition decreased versus increased performance on the Visual Detection Task. Interestingly, activity of ATN-projecting TRN neurons (to anterior thalamic nucleus) positively correlated with arousal, and negatively with spindle power. Neither activation nor their inhibition impacted performance. I hypothetised that anteromedial thalamic nucleus (AM) is important for PFC guided recall of specific events/memories from  retrosplenial cortex - serving as a dialing machine/index, thus active in wakefulness. During visual detection task with only one rule to follow and no recall to do, it would not be recruited.

Steriade et al. (1993) demonstrated that both SWS and anesthetics induce cortical slow waves: Under anesthesia, the slow oscillation around 0.3-0.4 Hz is stereotyped, and during its depolarised phase occur spindles at 7-14 Hz and delta oscillations at l-4 Hz. During natural sleep, the slow oscillation has the same components and temporal sequences as during anesthesia, but spindles and delta rhythms dominate the EEG activity in some epochs. Further, in 38% of their recorded cortical neurons, the slow rhythm was combined with delta oscillation 1-4 Hz (both were phase locked to the slow and delta oscillations in the EEG). Delta activity occurred as stereotyped, clock-like action potentials during the depolarization phase of the slow rhythm. In 26% of recorded neurons, the slow rhythm was combined with spindle oscillations. Regular-spiking cortical neurons fully reflected the whole frequency range of thalamically generated spindles (7-14 Hz). However, during similar patterns of EEG spindling, intrinsically bursting cells fired grouped action potentials (with intraburst frequencies of 100-200 Hz) at only 2-4 Hz. The slow rhythm survived extensive thalamic neurons lesions with inputs to the recorded cortical neurons (Steriade et al., 1993). My model proposes that MHb-IPN-MRN-claustrum induced cortical and striatal slow waves synchronize the thalamo-cortical (spindles) and cortico-striato-thalamo-cortical memory replay (probabilistic cognitive, affective and sensory-motor learning) during the depolarized phases of SWS. Besides synchronizing the hippocampo-cortical replay of sequentialy bound information during ripples.

My work proposed applicable neural mechanisms for SWS, loss of awareness, memory replay, SWA, anesthesia, OIRD, antidepressant/anxiolytic, and anesthetic effects of psychedelics and ketamine, stress-sleep-resilience interactions, and for placebo effects. It predicted specific MHb-IPN system  interactions with neurotransmitters, neuromodulators, memory, safety, and danger. This knowledge helps to specifically target human brain to improve mental health and well-being.

## Conclusion: Testable Predictions and Future Directions

Proposed neural mechanisms showed HOW activation of MHb by mu-opioids, nicotine, acetylcholine, ATP, glutamate, substance P, GLP-1RA, interleukin 18, and disinhibition of MHb by endocannabinoids activates the IPN which induces SWS and its repair, growth and protective roles. The IPN induces memory replay by projections to dentate gyrus, hDBB, and activates MRN which is known to inhibit the theta-producing MS/vDBB. The IPN and MRN mostly gabaergic terminals inhibit other theta- and arousal-  promoting regions. Serotonergic MRN terminals activate claustrum which is known to induce slow-wave activity in cortex and striatum.
My neural model adds the MHb→IPN→MRN circuit into the SWS and anesthesia picture. It generates several key, falsifiable predictions:

1. Optogenetic/Chemogenetic Studies: Selective activation of MOR-expressing neurons in the MHb will recapitulate OIRD and induce SWA, while their inhibition will attenuate opioid-induced sedation and respiratory depression. Specific MHb→ IPN→ PAG pathway will cause OIRD. The IPI part of the IPN might nhibit the rest of IPN.
2. Anesthetic Mechanisms: Inhibition of IPN and/or MRN will stop anesthesia induced by propofol or isoflurane. Microinjection of GABA A receptor antagonists into the vPAG will decrease the inhibitory effect of the IPN→vPAG projections, and delay sedation.
3. Ketamine's Site of Action: Local administration of a 5-HT2A receptor antagonist into the IPN will attenuate the anesthetic, SWA, antidepressant, and withdrawal suppressing effects of 5-HT2a and ketamine (but ketamine is addictive in wakefulness, via VTA dopamine).
4. Circuit-Specificity: Inhibition of the claustrum will disrupt cortical SWA induced by MHb-IPN activation or ketamine administration, confirming its role as a critical downstream effector for loss of awareness.

5. Resilience mechanism: MHb, IPN, and MRN are mutually inhibiting the LHb, some of them also the MS/vDBB, SUM, VTA, LC, LH, PH, LDT, PPT, DRN, SI, vPAG, PBN regions. Thus functional connectivity MRI (fc MRI) and seed based analyses will confirm their negative coupling (temporal anticorrelation of their activities). This will prove that SWS has antidepressant, anxiolytic, and resilience to stress promoting effects. Model predicted that stress mutually inhibit MHb-IPN-MRN. MHb-IPN-MRN activity will be positively coupled with claustrum in fcMRI analyses. The LHb seed region will show anticorrelations with the MHb-IPN-MRN-claustrum circuit.

Testing these predictions will refine, or falsify the proposed mechanisms. Their translation will lead to more targeted and safer therapeutic strategies. This model suggests several Clinical and Translational Aplications:
1. OIRD Prevention: Selective inhibition of the MHb-IPN-PAG suppression of respiratory centers while preserving analgesic pathways
2. Slowing Neurodegeneration: MHb-SWS-serotonin-linked model of repair and neuroprotection
3. Sleep Therapeutics: MHb/IPN-targeted interventions for insomnia that preserve natural sleep architecture and anti-inflammatory roles
4. Depression Treatment: Circuit-informed 5-HT2a or ketamine protocols that maximize antidepressant efficacy, and decrease addiction. Application of MHb-SWS-resilience mechanism

5. Epilepsy: Anesthetics suppress hippocampal theta and breathing. 5-HT2a causes SWA + ripples

The critical challenge remains achieving sufficient spatial specificity for safe clinical manipulation of this circuit. Activation/inhibition of specific MHb-IPN-MRN→ claustrum + hippocampus + hDBB + SI/NBM + MS/vDBB + cortex circuits, their receptors and projections, will test their predicted roles. Translation of model’s findings into targeted neuropharmacological research will lead to safer treatments in depression, anxiety, trauma, autism, addiction, sleep, memory, epilepsy, neural repair, inflammation, and anesthesia.

Declaration of interest: none.